\documentclass[12pt,a4paper]{article}

\usepackage{amsthm, amsmath, amssymb}
\usepackage{amsmath}
\usepackage[table]{xcolor}
\usepackage{colortbl}
\usepackage{multirow}
\usepackage{bm}
\usepackage{graphicx}
\usepackage{booktabs}
\usepackage{adjustbox}
\usepackage{diagbox}
\usepackage{rotating}
\usepackage[margin = 2.5cm]{geometry}
\usepackage{setspace}

\usepackage[backend=bibtex, natbib=true, style=authoryear, maxnames=99]{biblatex}
\bibliography{ref.bib}
\AtEveryBibitem{%
  \clearlist{publisher}%
  \clearfield{issn}%
}

\renewcommand{\cite}{\citet}

\graphicspath{{img/}}

\begin{document}

\title{\textbf{Sparse Bayesian Vector Autoregressions in Huge Dimensions}}

\author{\textsc{Gregor Kastner}\\
Institute for Statistics and Mathematics\\
%Department of Finance, Accounting and Statistics\\
WU Vienna University of Economics and Business, Austria\\
\href{mailto:gregor.kastner@wu.ac.at}{gregor.kastner@wu.ac.at}
\\[.5em]
and \\[.5em]
\textsc{Florian Huber}\\
Salzburg Centre of European Union Studies (SCEUS)\\
University of Salzburg, Austria\\
\href{mailto:florian.huber@sbg.ac.at}{florian.huber@sbg.ac.at}}

\maketitle

\onehalfspacing

\begin{abstract}
\noindent We develop a Bayesian vector autoregressive (VAR) model with multivariate stochastic volatility that is capable of handling vast dimensional information sets. Three features are introduced to permit reliable estimation of the model. First, we assume that the reduced-form errors in the VAR feature a factor stochastic volatility structure, allowing for conditional equation-by-equation estimation. Second, we apply recently developed global-local shrinkage priors to the VAR coefficients to cure the curse of dimensionality. Third, we utilize recent innovations to efficiently sample from high-dimensional multivariate Gaussian distributions. This makes simulation-based fully Bayesian inference feasible when the dimensionality is large but the time series length is moderate. We demonstrate the merits of our approach in an extensive simulation study and apply the model to US macroeconomic data to evaluate its forecasting capabilities.
\end{abstract}


\vspace{1em}

\noindent \textbf{Acknowledgments:} The authors acknowledge funding from the Austrian Science Fund (FWF) for the project ``High-dimensional statistical learning: New methods to advance economic and sustainability policies'' (ZK 35), jointly carried out by WU Vienna University of Economics and Business, Paris Lodron University Salzburg, TU Wien, and the Austrian Institute of Economic Research (WIFO).

%\vspace{1em}

%\noindent \textbf{MSC2010:} 62F15, 62P20, 65C05, 65C60

\pagebreak

\section{Introduction}
Previous research has identified two important features that macroeconometric models should possess: the ability to exploit high dimensional information sets \citep{banbura2010, stock2011dynamic, roc-mca:dyn, koop2019bayesian} and the possibility to capture non-linear features of the underlying time series \citep{Cogley2002, primiceri2005time,clark2011real, Clark2015, bitto2015achieving, huber2019should}. While the literature suggests several paths to estimate large models, the majority of such approaches implies that once non-linearities are taken into account, analytical solutions are no longer available and the computational burden becomes prohibitive.\footnote{Two recent exceptions are \cite{Koop2013185} and \cite{carriero2015common}.} This implies that high dimensional non-linear models can practically be estimated only under strong (and often unrealistic) restrictions on the dynamics of the model. However, especially in forecasting applications or in structural analysis, successful models should generally be able to exploit lots of information and also control for breaks in the autoregressive parameters or, more importantly, changes in the volatility of economic shocks \citep{primiceri2005time, Sims2006a, koop2009evolution}.

Two reasons limit the use of large (or even huge) non-linear models. The first reason is statistical. Since the number of parameters in a standard vector autoregression rises quadratically with the number of time series included and commonly used macroeconomic time series are rather short, in-sample overfitting turns out to be a serious issue. As a solution, the Bayesian literature on VAR modeling \citep[e.g.][]{Doan1984,Litterman1986,Sims1998, george2008bayesian, banbura2010,clark2011real,koop2013forecasting,Clark2015, korobilis2016, fol-yu:ach, huber2016adaptive, ankargren2019flexible} suggests shrinkage priors that push the parameter space towards some stylized prior model like a multivariate random walk. On the other hand, \cite{ahe-etal:spa} suggest to view VARs as  graphical models and perform model selection drawing from the literature on sparse directed acyclic graphs. This typically leads to much improved forecasting properties and more meaningful structural inference. Moreover, the majority of the literature on Bayesian VARs imposes conjugate priors on the autoregressive parameters, allowing for analytical posterior solutions and thus avoiding simulation based techniques like Markov chain Monte Carlo (MCMC). Frequentist approaches often consider multi-step approaches \citep[e.g.][]{dav-etal:spa}.

The second reason is computational. Non-linear Bayesian models typically have to be estimated by means of MCMC, and computational intensity increases vastly when the number of component series becomes large. This increase stems from the fact that standard algorithms for multivariate regression models call for the inversion of large covariance matrices. Especially for sizable systems, this can quickly turn prohibitive since the inverse of the posterior variance-covariance matrix on the coefficients has to be computed for each sweep of the MCMC algorithm. For natural conjugate models, this step can be vastly simplified because the likelihood possesses a convenient Kronecker structure, implying that all equations in the VAR feature the same set of explanatory variables. This speeds up computation by large margins but restricts the flexibility of the model. \cite{carriero2015common}, for instance, exploit this fact and introduce a simplified stochastic volatility specification. Another strand of the literature augment each equation of the VAR by including the residuals of the preceding equations \citep{carriero2019} which also provides significant improvements in terms of computational speed. Finally, in a recent contribution, \cite{koop2019bayesian} reduce the dimensionality of the problem at hand by randomly compressing the lagged endogenous variables in the VAR.

All papers mentioned hitherto focus on capturing cross-variable correlation in the conditional mean through the VAR part and the co-movement in volatilities is captured by a rich specification of the error variance \citep{primiceri2005time} or by a single factor \citep{carriero2015common}. Another strand of the literature, typically used in financial econometrics, utilizes factor models to provide a parsimonious representation of a covariance matrix, focusing exclusively on the second moment of the predictive density. For instance, \cite{pitt1999time} and \cite{aguilar2000bayesian} assume that the variance-covariance matrix of a broad panel of time series might be described by a lower dimensional matrix of latent factors featuring stochastic volatility and a variable-specific idiosyncratic stochastic volatility process. 

The present paper  combines the virtues of exploiting  large information sets and allowing for  movements in the error variance. The overfitting issue mentioned above is solved as follows. First, we use a Dirichlet-Laplace (DL) prior specification \citep[see][]{bhattacharya2015dirichlet} on the VAR coefficients. This prior is a global-local shrinkage prior in the spirit of \cite{polson2010shrink} that enables us to heavily shrink the parameter space but at the same time provides enough flexibility to allow for non-zero regression coefficients if necessary. Second, a factor stochastic volatility model on the VAR errors grants a parsimonious representation of the time-varying error variance-covariance matrix of the VAR. To deal with the computational complexity, we exploit the fact that, conditionally on the latent factors and their loadings, equation-by-equation estimation becomes possible within each MCMC iteration. Moreover, we apply recent advances for fast sampling from high dimensional multivariate Gaussian distributions \citep{bhattacharya2015fast} that permit estimation of models with hundreds of thousands of autoregressive parameters and an error covariance matrix with tens of thousands of nontrivial time-varying elements on a quarterly US dataset in a reasonable amount of time. In a careful analysis, we show to what extent our proposed method improves upon a set of standard algorithms typically used to simulate from the joint posterior distribution of large dimensional Bayesian VARs.

We first assess the merits of our approach in an extensive simulation study based on a range of different data generating processes. Relative to a set of competing benchmark specifications we show that, in terms of point estimates, the proposed global-local shrinkage prior yields precise parameter estimates and successfully introduces shrinkage in the modeling framework, without overshrinking significant signals.

In an empirical application, we adopt a modified version of the quarterly dataset proposed by \cite{stock2011dynamic} and \cite{mccracken2016fred}. To illustrate the out-of-sample performance of our model, we forecast important economic indicators such as output, consumer price inflation and short-term interest rates, amongst others. The proposed model is benchmarked against several alternatives. Our findings suggests that it performs well in terms of one-step-ahead predictive likelihoods. In addition, investigating the time profile of the cumulative log predictive likelihood reveals that allowing for large information sets in combination with the factor structure especially pays off in times of economic stress.

The remainder of this paper is structured as follows. Section~\ref{sec:framework} introduces the econometric framework. Section~\ref{sec:est} details the Bayesian estimation approach, including an elaborated account of the (shrinkage) prior setup adopted and the corresponding conditional posterior distributions. Section~\ref{sec:comp} provides an analysis of the computational gains of our algorithm relative to a set of established algorithms. Section~\ref{sec:simstudy} presents the results of an extensive simulation study comparing the performance of carefully selected shrinkage priors for different time series lengths and model dimensions within various (sparse and dense) data generating scenarios.
Section~\ref{sec:appli}, after giving a brief overview of the dataset used along with the model specification, illustrates our modeling approach by fitting a single factor model to 215-dimensional quarterly US data. Moreover, we perform a forecasting exercise to assess the predictive performance of our approach and discuss the choice of the number of latent factors. Finally, Section~\ref{sec:concl} concludes.

\section{Econometric framework}\label{sec:framework}
%This section outlines  the vector autoregressive model with factor stochastic volatility. After carefully introducing the model we discuss its most important features and implications of our framework and briefly describe the identification strategy adopted.
%\subsection{The vector autoregressive model with factor stochastic volatility}

Suppose interest centers on modeling an $m \times 1$ vector of time series denoted by $\boldsymbol{y}_t$ with $t=1,\dots,T$.  We assume that $\boldsymbol{y}_t$ follows a heteroscedastic VAR($p$) process,\footnote{For simplicity of exposition we omit the intercept term in the following discussion (which we nonetheless include in the empirical application).}
\begin{align}
\boldsymbol{y}_t &= \boldsymbol{A}_1 \boldsymbol{y}_{t-1}+\dots +\boldsymbol{A}_p \boldsymbol{y}_{t-p} + \boldsymbol{\varepsilon}_t, \quad \boldsymbol{\varepsilon}_t \sim \mathcal{N}_m(\boldsymbol{0}, \boldsymbol{\Omega}_t).
\label{eq:VARFSV}
\end{align}
Each $\boldsymbol{A}_j~(j=1,\dots,p)$ is an $m \times m$ matrix of autoregressive coefficients. The error term is assumed to follow a multivariate Gaussian distribution with time-varying variance-covariance matrix $\boldsymbol{\Omega}_t$. To permit reliable and parsimonious estimation when $m$ is large, we decompose the residual covariance matrix into
\begin{equation}
\boldsymbol{\Omega}_t = \boldsymbol{\Lambda}\boldsymbol{V}_t \boldsymbol{\Lambda} + \boldsymbol{\Sigma}_t,
\label{eq:omega}
\end{equation}
where both $\boldsymbol{\Sigma}_t = \text{diag}(\sigma^2_{1t},\dots,\sigma^2_{mt})$ and $\boldsymbol{V}_t = \text{diag}(e^{h_{1t}},\dots,e^{h_{qt}})$ are diagonal matrices with dimension $m$ and $q$, respectively, and $\boldsymbol{\Lambda}$ denotes an $m \times q$ matrix of factor loadings with typical element $\lambda_{ij}$ ($i=1,\dots,m$; $j=1,\dots,q$). The logarithms of the diagonal elements of $\boldsymbol{\Sigma}_t$ and $\boldsymbol{V}_t$ follow  AR($1$) processes,
\begin{align}
h_{jt} &= \rho_{hj} h_{j,t-1} +  e_{hj,t}, \quad j=1,\dots,q, \label{eq:stateH}\\
\log \sigma^2_{it} &=\mu_{\sigma i}+\rho_{\sigma i} (\log \sigma^2_{i,t-1}-\mu_{\sigma i}) +  e_{\sigma i,t}, \quad i=1,\dots,m. \label{eq:stateSIG}
\end{align}
To identify the scaling of the elements of $\boldsymbol{\Lambda}$, the process specified in (\ref{eq:stateH}) is assumed to have mean zero while $\mu_{\sigma j}$ in (\ref{eq:stateSIG}) is the unconditional mean of the log-elements of $\boldsymbol{\Sigma}_t$ to be estimated from the data \citep[cf.][]{kas-etal:eff}. The parameters $\rho_{hj}$ and $\rho_{\sigma i}$ are a priori restricted to the interval $(-1,1)$ and denote the persistences of the latent log variances. The error terms $e_{hj, t}$ and $e_{\sigma i,t}$ constitute independent zero mean innovations with variances $\varsigma^2_{hj}$ and $\varsigma^2_{\sigma i}$, respectively. This specification implies that the volatilities are mean reverting and thus bounded in the limit. 

This error structure is known as the factor stochastic volatility model \citep[see e.g.][]{pitt1999time, aguilar2000bayesian}. It can be equivalently written by introducing $q$ conditionally independent latent factors $\boldsymbol{f}_t \sim \mathcal{N}_q(\boldsymbol{0}, \boldsymbol{V}_t)$ and rewriting the error term in (\ref{eq:VARFSV}) as
\begin{equation}
\boldsymbol{\varepsilon}_t = \boldsymbol{\Lambda}\boldsymbol{f}_t + \boldsymbol{\eta}_t, \quad \boldsymbol{\eta}_t \sim \mathcal{N}_m(\boldsymbol{0}, \boldsymbol{\Sigma}_t).
\label{eq:statfac}
\end{equation}
%|\boldsymbol{\Lambda},\boldsymbol{f}_t,\boldsymbol{\eta}_t
Note that off-diagonal entries of $\boldsymbol{\Omega}_t$ exclusively stem from the volatilities of the $q$ factors while the diagonal entries of $\boldsymbol{\Omega}_t$ are allowed to feature idiosyncratic deviations driven by the elements of $\boldsymbol{\Sigma}_t$. This specification reduces the number of free elements in $\boldsymbol{\Omega}_t$ from $m(m+1)/2$ to $mq$, where the latter quantity is typically much smaller than the former. In addition, by conditioning on the latent factors, this representation enables us to derive an efficient Gibbs sampler that allows for conditional equation-by-equation estimation. As  will be discussed in more detail in Section~\ref{sec:posteriors}, this constitutes a key feature for computationally feasible Bayesian inference when the dimensionality $m$ becomes large.
%The assumption that $\boldsymbol{\Sigma}$ is constant throughout time is crucial to exploit a convenient Kronecker structure on the likelihood function. This implies that the computational burden is reduced considerably, since in principal the system could be estimated equation by equation. The statistical implications are that each variable in $\boldsymbol{y}_t$ features a constant long-run variance, whereas deviations from that steady state value tend to be driven by the latent factors. 

The model described by Eqs.\ (\ref{eq:VARFSV}) to (\ref{eq:omega}) is related to several alternative specifications commonly used in the literature. For instance, assuming that $\boldsymbol{V}_t = \boldsymbol{I}$ and $\boldsymbol{\Sigma}_t \equiv \boldsymbol{\Sigma}$ for all $t$ leads to the specification adopted in \cite{stock2005understanding}.  Setting $q=1$ and $\boldsymbol{\Sigma}_t \equiv \boldsymbol{\Sigma}$ yields a specification that is similar to the one stipulated in \cite{carriero2015common}, with the difference that our model imposes restrictions on the covariances whereas \cite{carriero2015common} estimate a full (but constant) covariance matrix. In addition, our model implies that the stochastic volatility enters $\boldsymbol{\Omega}_t$ in an additive fashion. 

Before proceeding to the next subsection it is worth summarizing the key features of the model given by Eqs.\ (\ref{eq:VARFSV}) to (\ref{eq:statfac}). First, we capture cross-variable movements in the conditional mean through the VAR block of the model and assume that co-movement in conditional variances is captured by a factor structure.
Second, the model introduces stochastic volatility by assuming that a large panel of volatilities may be efficiently summarized through a set of latent heteroscedastic factors. This choice is more flexible than a single factor model for the volatility, effectively providing a parsimonious representation of $\boldsymbol{\Omega}_t$ that is flexible enough to replicate the dynamic behavior of the variances of a broad set of macroeconomic quantities. 

\section{Inference in large dimensional VAR models}\label{sec:est}
Our approach to estimation and inference is Bayesian. This implies that after specifying a suitable prior distribution on the model parameters, we can combine this prior with the likelihood implied by the data and the model to obtain the corresponding posterior distribution. 
\subsection{A global-local shrinkage prior}
For prior implementation, it proves to be convenient to define a $k \times 1$ vector of predictors $\boldsymbol{x}_t =( \boldsymbol{y}'_{t-1},\dots,\boldsymbol{y}_{t-p}')'$  and an $m \times k$ coefficient matrix $\boldsymbol{B}=(\boldsymbol{A}_1,\dots,\boldsymbol{A}_p)$ with $k=mp$ to rewrite the model in (\ref{eq:VARFSV}) more compactly as
$
\boldsymbol{y}_t = \boldsymbol{B} \boldsymbol{x}_t + \boldsymbol{\varepsilon}_t.
$
Stacking the rows of $\boldsymbol{y}_t$, $\boldsymbol{x}_t$, and $\boldsymbol{\varepsilon_t}$ yields
\begin{equation}
\boldsymbol{Y} = \boldsymbol{X} \boldsymbol{B}' + \boldsymbol{E},
\end{equation}
where $\boldsymbol{Y} = (\boldsymbol{y}_1,\dots,\boldsymbol{y}_T)'$, $\boldsymbol{X} = (\boldsymbol{x}_1, \dots, \boldsymbol{x}_T)'$, and $\boldsymbol{E} = (\boldsymbol{\varepsilon}_1,\dots,\boldsymbol{\varepsilon}_T)'$ denote the corresponding full data matrices.

Typically, the matrix $\boldsymbol{B}$ is a sparse matrix  with non-zero elements mainly located on the main diagonal of  $\boldsymbol{A}_1$. In fact, existing priors in the Minnesota tradition tend to strongly push the  system towards  the prior model in high dimensions. However, especially in large models an extremely tight prior on $\boldsymbol{B}$ might lead to severe overshrinking, effectively zeroing out coefficients that might be important to explain $\boldsymbol{y}_t$. If the matrix $\boldsymbol{B}$ is characterized by a relatively low number of non-zero regression coefficients, a possible solution is a global-local shrinkage prior \citep{polson2010shrink}. 

A recent variant that falls within the class of global-local shrinkage priors is the Dirichlet-Laplace (DL) prior put forward in \cite{bhattacharya2015dirichlet}. This prior possesses convenient shrinkage properties in the presence of a large degree of sparsity of the parameter vector $\boldsymbol{b}=\text{vec}(\boldsymbol{B})$. In what follows, we impose the DL prior on each of the $K=mk$ elements of $\boldsymbol{b}$, denoted as $b_j$ for $j=1,\dots,K$,
\begin{equation}
b_j \sim \mathcal{DE}(\vartheta_j \zeta) \quad \Leftrightarrow \quad
b_j \sim \mathcal{N}(0, \psi_j \vartheta_j^2 \zeta^2), \quad \psi_j \sim \mathcal{E}(1/2), 
\end{equation}
where $\mathcal{DE}$ denotes the double exponential (Laplace) and $\mathcal{E}$ the exponential distribution, $\psi_j$ is an auxiliary scaling parameters to achieve conditional normality, and the elements of $\boldsymbol{\vartheta}=(\vartheta_1,\dots,\vartheta_K)'$ are local auxiliary scaling parameters that are bounded to the $(K-1)$-dimensional simplex $\mathcal{S}^{K-1}=\{\boldsymbol{\vartheta}: \vartheta_j \ge 0, \sum_{j=1}^n \vartheta_j =1\}$. A natural prior choice for $\vartheta_j$ is the (symmetric) Dirichlet distribution with hyperparameter $a$,
$
\vartheta_j\sim \mathcal{D}(a,\dots,a).
$
In addition, $\zeta$ is a global shrinkage parameter that pushes all elements in $\boldsymbol{B}$ towards zero and exhibits an important role in determining the tail behavior of the marginal prior distribution on $b_j$, obtained after integrating out the $\vartheta_j$s. Thus, we follow \cite{bhattacharya2015dirichlet} and adopt a fully Bayesian approach by specifying a gamma distributed prior on
$
 \zeta \sim \mathcal{G}(K a, 1/2).
$
It is noteworthy that this prior setup has at least two convenient features that appear to be of prime importance for VAR modeling. First, it exerts a strong degree of shrinkage on all elements of $\boldsymbol{B}$ but still provides additional flexibility such that non-zero regression coefficients are permitted. This critical property is a feature which a large class of global-local shrinkage priors share \citep{griffin2010inference,carvalho2010horseshoe,polson2010shrink} and has been recently adopted in a VAR framework by \cite{huber2016adaptive} and within the general context of state space models by \cite{bitto2015achieving}. Second, implementation is simple and requires relatively little additional input from the researcher. In fact, the prior heavily relies on a single structural hyperparameter that has to be specified with care, namely $a$. 

The hyperparameter $a$ influences the empirical properties of the proposed shrinkage prior along several important dimensions. Smaller values of $a$ lead to heavy shrinkage on all elements of $\boldsymbol{B}$. To see this, note that lower values of $a$ imply that more prior mass is placed on small values of $\zeta$ a priori. Similarly, when $a$ is small, the Dirichlet prior places more mass on values of $\vartheta_j$ close to zero. Since lower values of $\zeta$ translate into thicker tails of the marginal prior on $b_j$, the specific choice of $a$ not only influences the overall degree of shrinkage but also the tail behavior of the prior. Letting $\tilde p$ denote the number of predictors, \cite{bhattacharya2015dirichlet} show that if $a$ is specified as $\tilde p^{-(1+\Delta)}$ for any $\Delta>0$ to be small, the DL prior displays excellent posterior contraction rates, and \cite{pat-etal:pos} discuss the shrinkage properties of the proposed prior within the context of factor models. In our application, $\tilde p=K$ (when considering the total number of predictors) or $\tilde p=k$ (when considering the number or predictors per equation). 

For the factor loadings we independently use a standard normally distributed prior on each element $\lambda_{ij}\sim \mathcal{N}(0,1)$ for $i=1,\dots,m$ and $j=1,\dots,q$. In the empirical application (Section~\ref{sec:appli}), we in addition consider the row-wise normal-gamma \citep[NG,][]{griffin2010inference} shrinkage prior discussed in \cite{kas:spa}, i.e.\ $\lambda_{ij}|\tau_{ij}^2 \sim \mathcal{N}(0, \tau_{ij}^2)$, $\tau_{ij}^2|\upsilon_{i}^2 \sim \mathcal{G}(a_\lambda, a_\lambda \upsilon_{ij}^2/2)$, $\lambda_{ij}^2 \sim \mathcal{G}(c_\lambda, d_\lambda).$ Furthermore, we impose a normally distributed prior on the mean of the log-volatility $\mu_{\sigma j} \sim \mathcal{N}(0,M_\mu)$ with $M_\mu$ denoting the prior variance, and the commonly employed Beta distributed prior on the transformed persistence parameter of the log-volatility $\frac{\rho_{sj}+1}{2} \sim \mathcal{B}(a_0, b_0)$ for $s \in \{h, \sigma\}$ and $a_0, b_0 \in \mathbb{R}^+$ to ensure stationarity. Finally, we use a restricted gamma prior on the innovation variances in Eqs.\ (\ref{eq:stateH}) and (\ref{eq:stateSIG}), $\varsigma_{sj}^2  \sim \mathcal{G}(\frac{1}{2}, \frac{1}{2 \xi})$. Here, $\xi$ is a hyperparameter used to control the tightness of the prior. This choice, motivated in \cite{fruhwirth2010stochastic} implies that if the data is not informative on the degree of time variation of the log volatilities then we do not bound $\varsigma_{sj}^2$ artificially away from zero, effectively applying more shrinkage than the standard inverted gamma prior. 

\subsection{Full conditional posterior distributions}
\label{sec:posteriors}
Conditional on the  latent factors and the corresponding loadings,  the  model in (\ref{eq:VARFSV}) can be cast as a system of $m$ unrelated regression models for the elements in $\boldsymbol{z}_t =\boldsymbol{y}_t-\boldsymbol{\Lambda} \boldsymbol{f}_t$, labeled $z_{it}$, with heteroscedastic errors,
\begin{equation}
{z}_{it}= \boldsymbol{B}_{i \bullet} \boldsymbol{x}_t +\eta_{it}, \quad i=1,\dots,m. \label{eq: eqspecific}
\end{equation}
Here we let  $\boldsymbol{B}_{i \bullet}$ denote the  $i$th row of $\boldsymbol{B}$ and $\eta_{it}$ is the $i$th element of $\boldsymbol{\eta}_t$.  The corresponding posterior distribution of $\boldsymbol{B}'_{i\bullet}$ is $k$-variate Gaussian,
\begin{equation}
\boldsymbol{B}'_{i\bullet}|\bullet \sim \mathcal{N}(\boldsymbol{b}_i, \boldsymbol{Q}_i), \label{eq: postB}
\end{equation}
with $\bullet$ indicating that we condition on the remaining parameters and latent quantities of the model. The  posterior variance and mean are given by
\begin{align}
\boldsymbol{Q}_i &= (\tilde{\boldsymbol{X}}_i'\tilde{\boldsymbol{X}}_i+\boldsymbol{\Phi}_i^{-1})^{-1},\label{eq: VAR_B}\\ 
\boldsymbol{b}_i &= \boldsymbol{Q}_i  (\tilde{\boldsymbol{X}}_i' \tilde{\boldsymbol{z}_{i}}).\label{eq: meanB}
\end{align}
The diagonal prior covariance matrix  of the coefficients related to the $i$th equation  is given  by $\boldsymbol{\Phi}_i$, the respective $k \times k$ diagonal submatrix of  $\boldsymbol{\Phi}=\zeta \times \text{diag}(\psi_1 \vartheta_1^2 ,\dots,\psi_K \vartheta_K^2)$. Moreover, $\tilde{\boldsymbol{X}}_i$ is a $T\times k$ matrix with typical row $t$ given by  $\boldsymbol{X}_t/\sigma_{it}$ and $\tilde{\boldsymbol{z}}_{i}$ is a  $T$-dimensional vector with the $t$th element given by  $z_{it}/\sigma_{it}$. This normalization renders (\ref{eq: eqspecific}) conditionally homoscedastic with standard normally distributed white noise errors.

The full conditional posterior distribution of  $\psi_j$ is inverse Gaussian,
\begin{equation}
\psi_j|\bullet \sim {iG}(\vartheta_j \zeta/ |b_j|, 1), \quad j=1,\dots,K.
\end{equation}
The conditional posterior of the global shrinkage parameter $\zeta$ follows a generalized inverse Gaussian (GIG) distribution, \begin{equation}
\zeta|\bullet \sim \mathcal{GIG}\left(K(a - 1),1, 2\sum_{j=1}^K  |b_j|/\vartheta_j\right).
\end{equation}
To draw from this distribution, we use the efficient algorithm of \cite{hoe-ley:gen}. Moreover, we sample the  scaling parameters $\vartheta_j$   by first sampling  $L_j$ from
$
L_j |\bullet \sim \mathcal{GIG}(a-1, 1, 2|b_j|),
$
and then  setting 
$
\vartheta_j = L_j/\sum_{i=1}^K L_i.
$

The conditional posterior distributions of the factors are Gaussian and thus straightforward to draw from. The factor loadings are sampled using ``deep interweaving'' \citep[see][]{kas-etal:eff}, and the parameters in (\ref{eq:stateH}) and (\ref{eq:stateSIG}) along the full histories of the latent log-volatilities are sampled as in \cite{kastner2014ancillarity} using the
R-packages \texttt{factorstochvol} \citep{r:fac} and \texttt{stochvol} \citep{kas:dea}.

Our MCMC algorithm  iteratively draws from the conditional posterior distributions outlined above and discards the  first $J$ draws as burn-in. In terms of computational requirements, the single most intensive step is the simulation from  the joint posterior of the autoregressive coefficients in $\boldsymbol{B}$. Because this step is implemented on an equation-by-equation basis, speed improvements relative to  the standard approach are already quite substantial. However, note that if $k$ is large (i.e.\ of the order of several thousands), even the commonly employed equation-by-equation sampling fails to deliver a sufficient amount of draws within a reasonable time window.  Consequently, we  outline  an alternative  algorithm to draw from a high-dimensional multivariate Gaussian distribution under a  Bayesian prior that features a diagonal prior  variance-covariance matrix in the upcoming section.

\section{Computational aspects}\label{sec:comp}
The typical approach to sampling from (\ref{eq: postB}) is based on the full system and simultaneously samples from the full conditional posterior of $\boldsymbol{B}$, implying that the corresponding posterior distribution is a $K$-dimensional Gaussian distribution with a $K \times K$ dimensional variance-covariance matrix.  Under a non-conjugate prior, the computational difficulties arise from the need to invert the $K\times K$ variance-covariance matrix which requires operations of order  $O(m^6 p^3)$ under Gaussian elimination.

If a conjugate prior in combination with a constant \citep[or vastly simplified heteroscedastic, see][]{carriero2015common} specification of $\boldsymbol{\Omega}_t$ is used, the corresponding variance-covariance features a Kronecker structure which is computationally cheaper to invert and scales better in large dimensions. Specifically, the manipulations of the corresponding covariance matrix are of order $O(m^3+k^3)$, a significant gain relatively to the standard approach. However, this comes at a cost since all equations have to feature the same set of variables, the prior on the VAR coefficients has to be symmetric and any stochastic volatility specification that preserves conjugacy is necessary overly simplistic. 

By contrast, recent studies emphasize the computational gains that arise from utilizing a framework that is based on equation-by-equation estimation. \cite{carriero2019} and \cite{koop2019bayesian} augment each equation of the system by either contemporaneous  values of the endogenous variables of the preceding equations  or the residuals from the previous equations. Here, our approach renders the equations of the system conditionally independent by conditioning on the factors. From a computational perspective, the differences between using a factor model to disentangle the equations and an approach based on augmenting specific equations by quantities that aim to approximate covariance parameters are negligible. If we sample from (\ref{eq: postB}) directly, the computations involved are of order $O(m k^3)=O(m^4 p^3)$. This already poses significant improvements relative to full system estimation.

One contribution of the present paper is the application of the algorithm proposed by \cite{bhattacharya2015fast} and developed for univariate regression models under a global-local shrinkage prior.
%\footnote{The algorithm is not specific to global-local shrinkage priors but requires that the prior variance-covariance matrix is cheap to invert, i.e. diagonal.}
This algorithm is applied to each equation in the  system  and  cycles through the following steps:
\begin{enumerate}
\item Sample independently $\boldsymbol{u}_i \sim \mathcal{N}(\boldsymbol{0}_k, \boldsymbol{\Phi}_i)$ and  $\boldsymbol{\delta}_i \sim \mathcal{N}(\boldsymbol{0},\boldsymbol{I}_T)$.
\item  Use $\boldsymbol{u}_i$ and $\boldsymbol{\delta}_i$ to construct $\boldsymbol{v}_i =  \tilde{\boldsymbol{X}}_i \boldsymbol{u}_i+\boldsymbol{\delta}_i$.
\item  Solve  $(\tilde{\boldsymbol{X}}_i \boldsymbol{\Phi}_i  \tilde{\boldsymbol{X}}_i'+\boldsymbol{I}_T) \boldsymbol{w}_i=(\tilde{\boldsymbol{z}}_{i} -\boldsymbol{v}_i)$ for $\boldsymbol{w}_i$.
\item Set $\boldsymbol{B}'_{i \bullet}=\boldsymbol{u}_i+\boldsymbol{\Phi}_i \tilde{\boldsymbol{X}}_i'\boldsymbol{w}_i$.
\end{enumerate}
%\cite{bhattacharya2015fast} show that this algorithm yields a valid draw from  (\ref{eq: postB}).
This algorithm  outperforms all competing variants  discussed  previously in situations where $k \gg T$, a  situation  commonly encountered  when dealing with large VAR models.  In such cases,  steps (1) to (4) can be carried out using  $O(p m^2  T^2)$ floating point operations.  In situations where $k \approx T$, the computational advantages relative to  the  standard equation-by-equation algorithm mentioned above are modest or even negative. However, note that the cost is quadratic in $m$ and linear in $p$ and thus scales much better when the number of endogenous variables and/or lags thereof is increased. More information on the empirical performance of our algorithm can be found in Section~\ref{sec:comptime}.

\section{Simulation Study}
\label{sec:simstudy}

This section aims at comparing the performance of the DL prior with a range of commonly used alternatives. We investigate sparse, intermediate, and dense data generating processes (DGPs) where $T \in \{50, 100, 150, 200, 250\}$ and $m \in \{10, 20, 50, 100\}$. The probability of an off-diagonal entry to be non-zero is $0.01$, $0.1$, and $0.8$ in each of the respective scenarios. In all scenarios, each intercept entry has a $0.1$ probability of being non-zero and all diagonal elements are non-zero with probability $0.8$.
The non-zero elements are randomly generated from Gaussian distributions roughly tuned to yield stable VARs. More concretely, both the mean $\mu_I$ and the standard deviation $\sigma_I$ of the intercept are set to $0.01$, whereas mean and standard deviation of the diagonal ($D$) and the off-diagonal ($O$) elements are chosen as follows:
\begin{itemize}
\item Dense: 80\% off-diagonal density level, \(\mu_D = \sigma_D = 0.15\) and \(\mu_O = \sigma_O = 0.01\).
\item Intermediate: 10\% off-diagonal density level, \(\mu_D = \sigma_D = 0.15\) and \(\mu_O = \sigma_O = 0.1\).
\item Sparse: 1\% off-diagonal density level, \(\mu_D = \sigma_D = \mu_O = \sigma_O = 0.3\).
\end{itemize}
Concerning the errors, we use a single factor SV specification. The factor loadings are generated from $\mathcal{N}(0.001, 0.001^2)$ to roughly match the above scaling. The AR(1) processes driving the idiosyncratic log-variances are assumed to have mean $\mu_{\sigma i} = -12$ with persistences $\rho_{\sigma i}$ ranging from $0.85$ to $0.98$ and innovation standard deviations $\varsigma_{\sigma i}$ from $0.3$ to $0.1$. The process driving the factor log variance is assumed to be highly persistent with $\rho_{h1} = 0.99$ and $\varsigma_{h1} = 0.1$.

\begin{table}[tp]
\caption[]{\label{tab:simstudy}Median RMSEs stemming from 10 simulations per setting.}
\centering
\begin{adjustbox}{max width=\textwidth}
\begin{tabular}{lrrrrrrrrrrrr}
\toprule
\multicolumn{1}{c}{\multirow{ 2}{*}{\diagbox{$T$}{$m$}}} & \multicolumn{4}{c}{sparse} & \multicolumn{4}{c}{intermediate} & \multicolumn{4}{c}{dense} \\
\cmidrule(l{3pt}r{3pt}){2-5} \cmidrule(l{3pt}r{3pt}){6-9} \cmidrule(l{3pt}r{3pt}){10-13}
  & 10 & 20 & 50 & 100 & 10 & 20 & 50 & 100 & 10 & 20 & 50 & 100\\
\midrule
\addlinespace[0.3em]
\multicolumn{13}{l}{\textbf{DL ($\bm{a_{DL} = 1/2}$)}}\\
\hspace{1em}50 & \cellcolor[HTML]{FFBCBF}{0.079} & \cellcolor[HTML]{FFBABD}{0.081} & \cellcolor[HTML]{FFB5B8}{0.085} & \cellcolor[HTML]{FFB2B5}{0.088} & \cellcolor[HTML]{FFB8BB}{0.083} & \cellcolor[HTML]{FFB8BB}{0.084} & \cellcolor[HTML]{FFB5B8}{0.086} & \cellcolor[HTML]{FFB0B4}{0.089} & \cellcolor[HTML]{FFBFC1}{0.077} & \cellcolor[HTML]{FFB9BC}{0.082} & \cellcolor[HTML]{FFB3B6}{0.087} & \cellcolor[HTML]{FFAEB1}{0.091}\\
\hspace{1em}100 & \cellcolor[HTML]{FFD3D5}{0.056} & \cellcolor[HTML]{FFD4D6}{0.056} & \cellcolor[HTML]{FFD4D6}{0.056} & \cellcolor[HTML]{FFD2D4}{0.058} & \cellcolor[HTML]{FFD0D2}{0.060} & \cellcolor[HTML]{FFD0D2}{0.060} & \cellcolor[HTML]{FFD1D3}{0.059} & \cellcolor[HTML]{FFCFD1}{0.061} & \cellcolor[HTML]{FFD4D6}{0.056} & \cellcolor[HTML]{FFD4D6}{0.056} & \cellcolor[HTML]{FFD2D4}{0.058} & \cellcolor[HTML]{FFD0D2}{0.060}\\
\hspace{1em}150 & \cellcolor[HTML]{FFDFE1}{0.043} & \cellcolor[HTML]{FFDCDE}{0.047} & \cellcolor[HTML]{FFDEE0}{0.045} & \cellcolor[HTML]{FFDDDF}{0.046} & \cellcolor[HTML]{FFDDDF}{0.045} & \cellcolor[HTML]{FFDADB}{0.050} & \cellcolor[HTML]{FFDBDC}{0.048} & \cellcolor[HTML]{FFDCDD}{0.048} & \cellcolor[HTML]{FFDFE1}{0.044} & \cellcolor[HTML]{FFDBDC}{0.049} & \cellcolor[HTML]{FFDCDE}{0.047} & \cellcolor[HTML]{FFDBDC}{0.049}\\
\hspace{1em}200 & \cellcolor[HTML]{FFE2E3}{0.040} & \cellcolor[HTML]{FFE3E4}{0.040} & \cellcolor[HTML]{FFE4E5}{0.039} & \cellcolor[HTML]{FFE4E5}{0.038} & \cellcolor[HTML]{FFDFE1}{0.043} & \cellcolor[HTML]{FFE1E2}{0.042} & \cellcolor[HTML]{FFE1E2}{0.042} & \cellcolor[HTML]{FFE2E3}{0.041} & \cellcolor[HTML]{FFE2E3}{0.041} & \cellcolor[HTML]{FFE3E4}{0.040} & \cellcolor[HTML]{FFE2E3}{0.040} & \cellcolor[HTML]{FFE1E2}{0.041}\\
\hspace{1em}250 & \cellcolor[HTML]{FFE4E6}{0.038} & \cellcolor[HTML]{FFE7E8}{0.034} & \cellcolor[HTML]{FFE8E9}{0.034} & \cellcolor[HTML]{FFE8E9}{0.034} & \cellcolor[HTML]{FFE4E5}{0.039} & \cellcolor[HTML]{FFE4E6}{0.038} & \cellcolor[HTML]{FFE5E7}{0.037} & \cellcolor[HTML]{FFE5E7}{0.036} & \cellcolor[HTML]{FFE4E6}{0.038} & \cellcolor[HTML]{FFE6E7}{0.036} & \cellcolor[HTML]{FFE6E7}{0.036} & \cellcolor[HTML]{FFE5E7}{0.037}\\
\addlinespace[0.3em]
\multicolumn{13}{l}{\textbf{DL ($\bm{a_{DL} = 1/k}$)}}\\
\hspace{1em}50 & \cellcolor[HTML]{FFD4D6}{0.055} & \cellcolor[HTML]{FFE0E1}{0.043} & \cellcolor[HTML]{FFE5E7}{0.037} & \cellcolor[HTML]{FFE9EA}{0.033} & \cellcolor[HTML]{FFC8CA}{0.069} & \cellcolor[HTML]{FFD7D9}{0.053} & \cellcolor[HTML]{FFDBDC}{0.049} & \cellcolor[HTML]{FFDDDF}{0.046} & \cellcolor[HTML]{FFD3D5}{0.057} & \cellcolor[HTML]{FFE2E3}{0.041} & \cellcolor[HTML]{FFEBEC}{0.029} & \cellcolor[HTML]{FFEFF0}{0.026}\\
\hspace{1em}100 & \cellcolor[HTML]{FFE0E1}{0.042} & \cellcolor[HTML]{FFE7E8}{0.035} & \cellcolor[HTML]{FFEDEE}{0.028} & \cellcolor[HTML]{FFF0F1}{0.025} & \cellcolor[HTML]{FFDADB}{0.050} & \cellcolor[HTML]{FFDCDE}{0.047} & \cellcolor[HTML]{FFDFE1}{0.044} & \cellcolor[HTML]{FFE1E2}{0.041} & \cellcolor[HTML]{FFE1E2}{0.042} & \cellcolor[HTML]{FFE9EA}{0.033} & \cellcolor[HTML]{FFEFF0}{0.026} & \cellcolor[HTML]{FFF1F2}{0.023}\\
\hspace{1em}150 & \cellcolor[HTML]{FFE9EA}{0.032} & \cellcolor[HTML]{FFEAEB}{0.032} & \cellcolor[HTML]{FFF1F2}{0.023} & \cellcolor[HTML]{FFF4F5}{0.020} & \cellcolor[HTML]{FFE0E1}{0.042} & \cellcolor[HTML]{FFE3E4}{0.040} & \cellcolor[HTML]{FFE3E4}{0.040} & \cellcolor[HTML]{FFE4E6}{0.038} & \cellcolor[HTML]{FFE6E7}{0.036} & \cellcolor[HTML]{FFECED}{0.028} & \cellcolor[HTML]{FFF1F2}{0.023} & \cellcolor[HTML]{FFF2F3}{0.021}\\
\hspace{1em}200 & \cellcolor[HTML]{FFEBEC}{0.031} & \cellcolor[HTML]{FFEFF0}{0.026} & \cellcolor[HTML]{FFF4F5}{0.020} & \cellcolor[HTML]{FFF5F6}{0.017} & \cellcolor[HTML]{FFE2E3}{0.041} & \cellcolor[HTML]{FFE4E6}{0.038} & \cellcolor[HTML]{FFE6E7}{0.036} & \cellcolor[HTML]{FFE7E8}{0.034} & \cellcolor[HTML]{FFE9EA}{0.032} & \cellcolor[HTML]{FFEEEF}{0.027} & \cellcolor[HTML]{FFF2F3}{0.022} & \cellcolor[HTML]{FFF4F5}{0.020}\\
\hspace{1em}250 & \cellcolor[HTML]{FFEBEC}{0.030} & \cellcolor[HTML]{FFF2F3}{0.022} & \cellcolor[HTML]{FFF5F6}{0.018} & \cellcolor[HTML]{FFF8F9}{0.015} & \cellcolor[HTML]{FFE6E7}{0.036} & \cellcolor[HTML]{FFE7E8}{0.035} & \cellcolor[HTML]{FFE9EA}{0.033} & \cellcolor[HTML]{FFEAEB}{0.032} & \cellcolor[HTML]{FFEBEC}{0.029} & \cellcolor[HTML]{FFEFF0}{0.025} & \cellcolor[HTML]{FFF4F5}{0.020} & \cellcolor[HTML]{FFF5F5}{0.019}\\
\addlinespace[0.3em]
\multicolumn{13}{l}{\textbf{DL ($\bm{a_{DL} = 1/K}$)}}\\
\hspace{1em}50 & \cellcolor[HTML]{FFCED0}{0.062} & \cellcolor[HTML]{FFE0E1}{0.043} & \cellcolor[HTML]{FFE4E6}{0.038} & \cellcolor[HTML]{FFE8E9}{0.033} & \cellcolor[HTML]{FFC5C7}{0.072} & \cellcolor[HTML]{FFD5D7}{0.054} & \cellcolor[HTML]{FFDADB}{0.050} & \cellcolor[HTML]{FFDCDD}{0.047} & \cellcolor[HTML]{FFCDCF}{0.063} & \cellcolor[HTML]{FFE4E5}{0.038} & \cellcolor[HTML]{FFEDEE}{0.028} & \cellcolor[HTML]{FFEEEF}{0.027}\\
\hspace{1em}100 & \cellcolor[HTML]{FFDADB}{0.050} & \cellcolor[HTML]{FFE4E6}{0.038} & \cellcolor[HTML]{FFEDEE}{0.028} & \cellcolor[HTML]{FFF0F1}{0.025} & \cellcolor[HTML]{FFD1D3}{0.059} & \cellcolor[HTML]{FFDBDC}{0.049} & \cellcolor[HTML]{FFDDDF}{0.046} & \cellcolor[HTML]{FFE0E1}{0.043} & \cellcolor[HTML]{FFDCDE}{0.047} & \cellcolor[HTML]{FFE7E8}{0.035} & \cellcolor[HTML]{FFEFF0}{0.026} & \cellcolor[HTML]{FFF1F1}{0.024}\\
\hspace{1em}150 & \cellcolor[HTML]{FFDFE1}{0.044} & \cellcolor[HTML]{FFE9EA}{0.033} & \cellcolor[HTML]{FFEFF0}{0.025} & \cellcolor[HTML]{FFF3F4}{0.021} & \cellcolor[HTML]{FFD6D8}{0.054} & \cellcolor[HTML]{FFDEE0}{0.044} & \cellcolor[HTML]{FFE2E3}{0.041} & \cellcolor[HTML]{FFE4E5}{0.039} & \cellcolor[HTML]{FFE4E5}{0.038} & \cellcolor[HTML]{FFE9EA}{0.033} & \cellcolor[HTML]{FFF1F1}{0.024} & \cellcolor[HTML]{FFF2F3}{0.022}\\
\hspace{1em}200 & \cellcolor[HTML]{FFE3E4}{0.040} & \cellcolor[HTML]{FFEAEB}{0.031} & \cellcolor[HTML]{FFF2F3}{0.022} & \cellcolor[HTML]{FFF5F6}{0.018} & \cellcolor[HTML]{FFD9DA}{0.051} & \cellcolor[HTML]{FFE1E2}{0.041} & \cellcolor[HTML]{FFE4E5}{0.039} & \cellcolor[HTML]{FFE6E7}{0.036} & \cellcolor[HTML]{FFE4E5}{0.039} & \cellcolor[HTML]{FFEBEC}{0.029} & \cellcolor[HTML]{FFF1F2}{0.022} & \cellcolor[HTML]{FFF3F4}{0.021}\\
\hspace{1em}250 & \cellcolor[HTML]{FFE5E7}{0.037} & \cellcolor[HTML]{FFEFF0}{0.026} & \cellcolor[HTML]{FFF5F5}{0.019} & \cellcolor[HTML]{FFF7F8}{0.016} & \cellcolor[HTML]{FFE0E1}{0.043} & \cellcolor[HTML]{FFE4E5}{0.039} & \cellcolor[HTML]{FFE6E7}{0.036} & \cellcolor[HTML]{FFE9EA}{0.033} & \cellcolor[HTML]{FFE4E5}{0.039} & \cellcolor[HTML]{FFEEEF}{0.027} & \cellcolor[HTML]{FFF3F4}{0.020} & \cellcolor[HTML]{FFF3F4}{0.020}\\
\addlinespace[0.3em]
\multicolumn{13}{l}{\textbf{NG ($\bm{a_{NG} = 1}$)}}\\
\hspace{1em}50 & \cellcolor[HTML]{FFCDCF}{0.063} & \cellcolor[HTML]{FFDBDC}{0.049} & \cellcolor[HTML]{FFDEE0}{0.044} & \cellcolor[HTML]{FFE1E2}{0.042} & \cellcolor[HTML]{FFC7C9}{0.069} & \cellcolor[HTML]{FFD6D8}{0.053} & \cellcolor[HTML]{FFDBDC}{0.049} & \cellcolor[HTML]{FFDCDE}{0.047} & \cellcolor[HTML]{FFD0D2}{0.060} & \cellcolor[HTML]{FFE1E2}{0.042} & \cellcolor[HTML]{FFEBEC}{0.031} & \cellcolor[HTML]{FFEEEF}{0.027}\\
\hspace{1em}100 & \cellcolor[HTML]{FFD9DA}{0.051} & \cellcolor[HTML]{FFE0E1}{0.042} & \cellcolor[HTML]{FFE5E7}{0.036} & \cellcolor[HTML]{FFE9EA}{0.033} & \cellcolor[HTML]{FFD6D8}{0.054} & \cellcolor[HTML]{FFDCDD}{0.048} & \cellcolor[HTML]{FFE0E1}{0.042} & \cellcolor[HTML]{FFE3E4}{0.039} & \cellcolor[HTML]{FFDCDD}{0.048} & \cellcolor[HTML]{FFE5E7}{0.037} & \cellcolor[HTML]{FFEDEE}{0.027} & \cellcolor[HTML]{FFF1F2}{0.023}\\
\hspace{1em}150 & \cellcolor[HTML]{FFE0E1}{0.043} & \cellcolor[HTML]{FFE5E7}{0.037} & \cellcolor[HTML]{FFEBEC}{0.031} & \cellcolor[HTML]{FFEDEE}{0.028} & \cellcolor[HTML]{FFDDDF}{0.046} & \cellcolor[HTML]{FFE0E1}{0.042} & \cellcolor[HTML]{FFE4E6}{0.038} & \cellcolor[HTML]{FFE6E7}{0.036} & \cellcolor[HTML]{FFE2E3}{0.040} & \cellcolor[HTML]{FFE7E8}{0.034} & \cellcolor[HTML]{FFEFF0}{0.026} & \cellcolor[HTML]{FFF2F3}{0.021}\\
\hspace{1em}200 & \cellcolor[HTML]{FFE4E5}{0.039} & \cellcolor[HTML]{FFE9EA}{0.033} & \cellcolor[HTML]{FFEDEE}{0.028} & \cellcolor[HTML]{FFF0F1}{0.025} & \cellcolor[HTML]{FFDFE1}{0.044} & \cellcolor[HTML]{FFE4E5}{0.039} & \cellcolor[HTML]{FFE6E7}{0.036} & \cellcolor[HTML]{FFE9EA}{0.032} & \cellcolor[HTML]{FFE4E6}{0.038} & \cellcolor[HTML]{FFEAEB}{0.031} & \cellcolor[HTML]{FFF0F1}{0.024} & \cellcolor[HTML]{FFF3F4}{0.020}\\
\hspace{1em}250 & \cellcolor[HTML]{FFE4E6}{0.037} & \cellcolor[HTML]{FFEBEC}{0.029} & \cellcolor[HTML]{FFF0F1}{0.025} & \cellcolor[HTML]{FFF1F2}{0.023} & \cellcolor[HTML]{FFE4E5}{0.039} & \cellcolor[HTML]{FFE6E7}{0.035} & \cellcolor[HTML]{FFE9EA}{0.032} & \cellcolor[HTML]{FFEBEC}{0.031} & \cellcolor[HTML]{FFE4E6}{0.038} & \cellcolor[HTML]{FFECED}{0.029} & \cellcolor[HTML]{FFF1F2}{0.022} & \cellcolor[HTML]{FFF5F5}{0.019}\\
\addlinespace[0.3em]
\multicolumn{13}{l}{\textbf{NG ($\bm{a_{NG} = 0.1}$)}}\\
\hspace{1em}50 & \cellcolor[HTML]{FFD2D4}{0.058} & \cellcolor[HTML]{FFE0E1}{0.043} & \cellcolor[HTML]{FFE4E6}{0.038} & \cellcolor[HTML]{FFE8E9}{0.034} & \cellcolor[HTML]{FFCACC}{0.066} & \cellcolor[HTML]{FFD8DA}{0.052} & \cellcolor[HTML]{FFDCDD}{0.048} & \cellcolor[HTML]{FFDEE0}{0.045} & \cellcolor[HTML]{FFD5D7}{0.055} & \cellcolor[HTML]{FFE1E2}{0.042} & \cellcolor[HTML]{FFEBEC}{0.029} & \cellcolor[HTML]{FFEEEF}{0.026}\\
\hspace{1em}100 & \cellcolor[HTML]{FFE0E1}{0.043} & \cellcolor[HTML]{FFE7E8}{0.035} & \cellcolor[HTML]{FFEDEE}{0.028} & \cellcolor[HTML]{FFF1F1}{0.024} & \cellcolor[HTML]{FFDADB}{0.050} & \cellcolor[HTML]{FFDEE0}{0.045} & \cellcolor[HTML]{FFE4E5}{0.039} & \cellcolor[HTML]{FFE5E7}{0.037} & \cellcolor[HTML]{FFDFE1}{0.044} & \cellcolor[HTML]{FFE8E9}{0.034} & \cellcolor[HTML]{FFEFF0}{0.026} & \cellcolor[HTML]{FFF2F3}{0.022}\\
\hspace{1em}150 & \cellcolor[HTML]{FFE7E8}{0.035} & \cellcolor[HTML]{FFEBEC}{0.031} & \cellcolor[HTML]{FFF1F1}{0.024} & \cellcolor[HTML]{FFF4F5}{0.020} & \cellcolor[HTML]{FFE3E4}{0.040} & \cellcolor[HTML]{FFE4E5}{0.039} & \cellcolor[HTML]{FFE8E9}{0.034} & \cellcolor[HTML]{FFEAEB}{0.032} & \cellcolor[HTML]{FFE5E7}{0.037} & \cellcolor[HTML]{FFEBEC}{0.031} & \cellcolor[HTML]{FFF1F2}{0.023} & \cellcolor[HTML]{FFF3F4}{0.020}\\
\hspace{1em}200 & \cellcolor[HTML]{FFEBEC}{0.031} & \cellcolor[HTML]{FFEEEF}{0.026} & \cellcolor[HTML]{FFF3F4}{0.021} & \cellcolor[HTML]{FFF5F6}{0.018} & \cellcolor[HTML]{FFE4E5}{0.038} & \cellcolor[HTML]{FFE8E9}{0.034} & \cellcolor[HTML]{FFEAEB}{0.031} & \cellcolor[HTML]{FFECED}{0.028} & \cellcolor[HTML]{FFEBEC}{0.031} & \cellcolor[HTML]{FFEDEE}{0.027} & \cellcolor[HTML]{FFF1F2}{0.022} & \cellcolor[HTML]{FFF5F5}{0.019}\\
\hspace{1em}250 & \cellcolor[HTML]{FFEBEC}{0.030} & \cellcolor[HTML]{FFF1F2}{0.023} & \cellcolor[HTML]{FFF5F5}{0.018} & \cellcolor[HTML]{FFF7F8}{0.016} & \cellcolor[HTML]{FFE8E9}{0.034} & \cellcolor[HTML]{FFEBEC}{0.030} & \cellcolor[HTML]{FFEDEE}{0.028} & \cellcolor[HTML]{FFEFF0}{0.026} & \cellcolor[HTML]{FFEBEC}{0.031} & \cellcolor[HTML]{FFF1F1}{0.024} & \cellcolor[HTML]{FFF4F5}{0.020} & \cellcolor[HTML]{FFF5F6}{0.018}\\
\addlinespace[0.3em]
\multicolumn{13}{l}{\textbf{Minnesota ($\bm{a_{M} = 0.001}$)}}\\
\hspace{1em}50 & \cellcolor[HTML]{FB8A8F}{0.135} & \cellcolor[HTML]{FB8A8F}{0.137} & \cellcolor[HTML]{FB8A8F}{0.164} & \cellcolor[HTML]{FF989D}{0.105} & \cellcolor[HTML]{FB8A8F}{0.134} & \cellcolor[HTML]{FB8A8F}{0.140} & \cellcolor[HTML]{FB8A8F}{0.153} & \cellcolor[HTML]{FF9EA2}{0.102} & \cellcolor[HTML]{FB8A8F}{0.131} & \cellcolor[HTML]{FB8A8F}{0.143} & \cellcolor[HTML]{FB8A8F}{0.165} & \cellcolor[HTML]{FF969B}{0.107}\\
\hspace{1em}100 & \cellcolor[HTML]{FFAEB1}{0.092} & \cellcolor[HTML]{FF9A9E}{0.105} & \cellcolor[HTML]{FB8A8F}{0.112} & \cellcolor[HTML]{FB8A8F}{0.136} & \cellcolor[HTML]{FFAAAD}{0.094} & \cellcolor[HTML]{FF9CA0}{0.104} & \cellcolor[HTML]{FD8F93}{0.109} & \cellcolor[HTML]{FB8A8F}{0.119} & \cellcolor[HTML]{FFAAAD}{0.094} & \cellcolor[HTML]{FF9A9E}{0.105} & \cellcolor[HTML]{FB8A8F}{0.116} & \cellcolor[HTML]{FB8A8F}{0.129}\\
\hspace{1em}150 & \cellcolor[HTML]{FFBCBF}{0.079} & \cellcolor[HTML]{FFB8BB}{0.083} & \cellcolor[HTML]{FFB3B6}{0.088} & \cellcolor[HTML]{FF9FA3}{0.102} & \cellcolor[HTML]{FFC0C2}{0.077} & \cellcolor[HTML]{FFB9BC}{0.082} & \cellcolor[HTML]{FFB4B7}{0.086} & \cellcolor[HTML]{FFABAF}{0.094} & \cellcolor[HTML]{FFBBBE}{0.080} & \cellcolor[HTML]{FFB8BB}{0.084} & \cellcolor[HTML]{FFAFB2}{0.090} & \cellcolor[HTML]{FF9FA3}{0.102}\\
\hspace{1em}200 & \cellcolor[HTML]{FFC6C8}{0.070} & \cellcolor[HTML]{FFC5C7}{0.071} & \cellcolor[HTML]{FFC1C3}{0.075} & \cellcolor[HTML]{FFBABD}{0.082} & \cellcolor[HTML]{FFC7C9}{0.069} & \cellcolor[HTML]{FFC6C8}{0.070} & \cellcolor[HTML]{FFC3C5}{0.073} & \cellcolor[HTML]{FFBFC1}{0.077} & \cellcolor[HTML]{FFC6C8}{0.070} & \cellcolor[HTML]{FFC4C6}{0.072} & \cellcolor[HTML]{FFBFC1}{0.077} & \cellcolor[HTML]{FFB5B8}{0.085}\\
\hspace{1em}250 & \cellcolor[HTML]{FFD2D4}{0.058} & \cellcolor[HTML]{FFCED0}{0.063} & \cellcolor[HTML]{FFCBCD}{0.065} & \cellcolor[HTML]{FFC6C8}{0.070} & \cellcolor[HTML]{FFD1D3}{0.059} & \cellcolor[HTML]{FFCDCF}{0.063} & \cellcolor[HTML]{FFCCCE}{0.064} & \cellcolor[HTML]{FFC9CB}{0.067} & \cellcolor[HTML]{FFD0D2}{0.060} & \cellcolor[HTML]{FFCCCE}{0.064} & \cellcolor[HTML]{FFC8CA}{0.068} & \cellcolor[HTML]{FFC5C7}{0.071}\\
\addlinespace[0.3em]
\multicolumn{13}{l}{\textbf{Minnesota ($\bm{a_{M} = 0.0001}$)}}\\
\hspace{1em}50 & \cellcolor[HTML]{FFC9CB}{0.067} & \cellcolor[HTML]{FFD8DA}{0.052} & \cellcolor[HTML]{FFDCDD}{0.048} & \cellcolor[HTML]{FFDDDF}{0.046} & \cellcolor[HTML]{FFC3C5}{0.073} & \cellcolor[HTML]{FFD6D8}{0.054} & \cellcolor[HTML]{FFDBDC}{0.049} & \cellcolor[HTML]{FFDEE0}{0.045} & \cellcolor[HTML]{FFCED0}{0.062} & \cellcolor[HTML]{FFE4E6}{0.037} & \cellcolor[HTML]{FFECED}{0.028} & \cellcolor[HTML]{FFF2F3}{0.022}\\
\hspace{1em}100 & \cellcolor[HTML]{FFCDCF}{0.063} & \cellcolor[HTML]{FFDADB}{0.050} & \cellcolor[HTML]{FFDCDE}{0.047} & \cellcolor[HTML]{FFDEE0}{0.045} & \cellcolor[HTML]{FFC8CA}{0.069} & \cellcolor[HTML]{FFD8DA}{0.052} & \cellcolor[HTML]{FFDCDE}{0.047} & \cellcolor[HTML]{FFDFE1}{0.044} & \cellcolor[HTML]{FFD1D3}{0.059} & \cellcolor[HTML]{FFE6E7}{0.036} & \cellcolor[HTML]{FFEDEE}{0.028} & \cellcolor[HTML]{FFF2F3}{0.022}\\
\hspace{1em}150 & \cellcolor[HTML]{FFCFD1}{0.061} & \cellcolor[HTML]{FFDBDC}{0.049} & \cellcolor[HTML]{FFDEE0}{0.045} & \cellcolor[HTML]{FFDFE1}{0.044} & \cellcolor[HTML]{FFCACC}{0.066} & \cellcolor[HTML]{FFDBDC}{0.049} & \cellcolor[HTML]{FFDEE0}{0.045} & \cellcolor[HTML]{FFE1E2}{0.042} & \cellcolor[HTML]{FFD4D6}{0.056} & \cellcolor[HTML]{FFE7E8}{0.035} & \cellcolor[HTML]{FFEDEE}{0.028} & \cellcolor[HTML]{FFF2F3}{0.021}\\
\hspace{1em}200 & \cellcolor[HTML]{FFCFD1}{0.061} & \cellcolor[HTML]{FFDCDD}{0.047} & \cellcolor[HTML]{FFDFE1}{0.044} & \cellcolor[HTML]{FFE0E1}{0.043} & \cellcolor[HTML]{FFCBCD}{0.065} & \cellcolor[HTML]{FFDCDD}{0.048} & \cellcolor[HTML]{FFDFE1}{0.044} & \cellcolor[HTML]{FFE2E3}{0.041} & \cellcolor[HTML]{FFD6D8}{0.054} & \cellcolor[HTML]{FFE8E9}{0.034} & \cellcolor[HTML]{FFEEEF}{0.027} & \cellcolor[HTML]{FFF2F3}{0.022}\\
\hspace{1em}250 & \cellcolor[HTML]{FFD2D4}{0.058} & \cellcolor[HTML]{FFDDDF}{0.046} & \cellcolor[HTML]{FFE1E2}{0.042} & \cellcolor[HTML]{FFE1E2}{0.041} & \cellcolor[HTML]{FFD1D3}{0.059} & \cellcolor[HTML]{FFDDDF}{0.046} & \cellcolor[HTML]{FFE1E2}{0.042} & \cellcolor[HTML]{FFE3E4}{0.040} & \cellcolor[HTML]{FFD7D9}{0.053} & \cellcolor[HTML]{FFE8E9}{0.033} & \cellcolor[HTML]{FFEEEF}{0.027} & \cellcolor[HTML]{FFF3F4}{0.021}\\
\addlinespace[0.3em]
\multicolumn{13}{l}{\textbf{OLS (if exists)}}\\
\hspace{1em}50 & \cellcolor[HTML]{FB8A8F}{0.158} & \cellcolor[HTML]{FB8A8F}{0.205} & \cellcolor[HTML]{E6E6E6}{  DNE} & \cellcolor[HTML]{E6E6E6}{  DNE} & \cellcolor[HTML]{FB8A8F}{0.158} & \cellcolor[HTML]{FB8A8F}{0.211} & \cellcolor[HTML]{E6E6E6}{  DNE} & \cellcolor[HTML]{E6E6E6}{  DNE} & \cellcolor[HTML]{FB8A8F}{0.160} & \cellcolor[HTML]{FB8A8F}{0.211} & \cellcolor[HTML]{E6E6E6}{  DNE} & \cellcolor[HTML]{E6E6E6}{  DNE}\\
\hspace{1em}100 & \cellcolor[HTML]{FF969B}{0.106} & \cellcolor[HTML]{FB8A8F}{0.128} & \cellcolor[HTML]{FB8A8F}{0.163} & \cellcolor[HTML]{E6E6E6}{  DNE} & \cellcolor[HTML]{FE9499}{0.107} & \cellcolor[HTML]{FB8A8F}{0.126} & \cellcolor[HTML]{FB8A8F}{0.157} & \cellcolor[HTML]{E6E6E6}{  DNE} & \cellcolor[HTML]{FD8F93}{0.110} & \cellcolor[HTML]{FB8A8F}{0.128} & \cellcolor[HTML]{FB8A8F}{0.165} & \cellcolor[HTML]{E6E6E6}{  DNE}\\
\hspace{1em}150 & \cellcolor[HTML]{FFB2B5}{0.088} & \cellcolor[HTML]{FFA4A8}{0.099} & \cellcolor[HTML]{FB8A8F}{0.112} & \cellcolor[HTML]{FB8A8F}{0.155} & \cellcolor[HTML]{FFB3B6}{0.087} & \cellcolor[HTML]{FFA6A9}{0.098} & \cellcolor[HTML]{FD8F93}{0.109} & \cellcolor[HTML]{FB8A8F}{0.155} & \cellcolor[HTML]{FFAFB2}{0.090} & \cellcolor[HTML]{FFA1A5}{0.101} & \cellcolor[HTML]{FB8A8F}{0.115} & \cellcolor[HTML]{FB8A8F}{0.167}\\
\hspace{1em}200 & \cellcolor[HTML]{FFBBBE}{0.080} & \cellcolor[HTML]{FFBDC0}{0.078} & \cellcolor[HTML]{FFB3B6}{0.087} & \cellcolor[HTML]{FF9A9E}{0.104} & \cellcolor[HTML]{FFBCBF}{0.079} & \cellcolor[HTML]{FFBDC0}{0.078} & \cellcolor[HTML]{FFB5B8}{0.085} & \cellcolor[HTML]{FF989D}{0.106} & \cellcolor[HTML]{FFBBBE}{0.080} & \cellcolor[HTML]{FFBCBF}{0.080} & \cellcolor[HTML]{FFAEB1}{0.091} & \cellcolor[HTML]{FB8A8F}{0.114}\\
\hspace{1em}250 & \cellcolor[HTML]{FFCBCD}{0.065} & \cellcolor[HTML]{FFC5C7}{0.071} & \cellcolor[HTML]{FFBDC0}{0.078} & \cellcolor[HTML]{FFAEB1}{0.092} & \cellcolor[HTML]{FFCACC}{0.066} & \cellcolor[HTML]{FFC6C8}{0.070} & \cellcolor[HTML]{FFBFC1}{0.077} & \cellcolor[HTML]{FFB5B8}{0.086} & \cellcolor[HTML]{FFC9CB}{0.067} & \cellcolor[HTML]{FFC5C7}{0.071} & \cellcolor[HTML]{FFBABD}{0.081} & \cellcolor[HTML]{FFA7AB}{0.096}\\
\bottomrule
\end{tabular}
\end{adjustbox}
\end{table}

For each of the $60$ settings, we simulate $10$ data sets. For each of these, we run our MCMC algorithm to obtain $2000$ posterior draws after a burn-in of $1000$. Consequently, the posterior means are compared to the true values and root mean squared errors are computed. Finally, the median of each of these is reported in Table~\ref{tab:simstudy}. Alongside the DL prior with weak ($a_{DL} = a = 1/2$) and strong ($a_{DL} = a = 1/k$ and $a_{DL} = a = 1/K$) shrinkage, we also consider the NG prior with a single global shrinkage parameter \citep[see][for the exact specification]{huber2016adaptive} and a standard conjugate Minnesota prior with a single shrinkage parameter $a_M$, implemented by using dummy observations. For the NG prior we specify the prior on the global shrinkage parameter to induce heavy shrinkage (by setting both hyperparameters of the gamma prior equal to $0.01$) and the prior controlling the excess kurtosis $a_{NG}$ is set equal to $1$, corresponding to the Bayesian Lasso  \citep[see][]{par-cas:bay}, and $a_{NG}=0.1$. The latter choice places significant prior mass around zero but at the same time leads to a heavy tailed marginal prior. Finally, we report RMSEs of the OLS estimator (if it exists).

As is to be expected, Table~\ref{tab:simstudy} reveals strong to severe overfitting of OLS (corresponding to the posterior mode under a flat prior) which can be mitigated to a certain extent when the Minnesota prior with $a_M = 0.001$ is employed instead. Similarly, the DL prior with weak shrinkage ($a_{DL} = 1/2$) displays a tendency to overfit, in particular when $T$ is small.
By contrast, the more aggressive DL and NG shrinkage priors show superior performance.
Overall, DL($1/k$) and NG($0.1$) exhibit lowest RMSEs, where DL($1/k$) performs best in the sparse scenarios, NG($0.1$) performs best in the intermediate settings, and no clear winner is to be found in the dense context.
Turning towards NG($1$) and DL($1/K$) we tend to observe acceptable but slightly inferior overall performance.
The Minnesota prior with $a_M = 0.0001$  yields an extreme degree of shrinkage, translating into estimates of autoregressive coefficients that are very close to zero, irrespectively of the contribution from the likelihood. In that sense, it overshrinks most of the nonzero coefficients. Nevertheless, in scenarios with extremely low signal-to-noise ratios (such as the dense scenario with $T=50$ and $m=100$), this can be beneficial for the overall performance.

For further illustration, we showcase four exemplary scenarios in Figures~\ref{simstudyexample1} to \ref{simstudyexample4} in the Appendix.

\section{Empirical forecasting application}\label{sec:appli}
In Section~\ref{sec:data} we first summarize the data set adopted and present the model specification choices made. The section that follows (Section~\ref{sec:results}) estimates a simple one factor model to outline the virtues of our proposed framework. Section~\ref{sec:forecast} presents the main findings of our forecasting exercise and discusses the choice of the number of factors used for modeling the error covariance structure.

\subsection{Data, model specification and selection issues} \label{sec:data}
Aim of the empirical application is to forecast a set of key US macroeconomic quantities. To this end, we use the quarterly dataset provided by \cite{mccracken2016fred}, a variant of the well-known \cite{stock2011dynamic} dataset for the US.\footnote{In addition to quarterly observations, \cite{mccracken2016fred} also provide a subset of the data which is observed monthly. Of course, our method is analogously applicable to higher frequency observations. However, given that the computational cost of the \cite{bhattacharya2015fast} approach is quadratic in $T$, the run-time gains of their approach in comparison to equation-by-equation estimation is then smaller and can, depending on the number of lags, even become be negative.}
The data spans the period ranging from 1959:Q1 to 2015:Q4. We include $m=215$ quarterly time series, capturing information on $14$ important segments of the economy and follow \cite{mccracken2016fred} in transforming the data to be approximately stationary.
Furthermore, we standardize each component series to have zero mean and variance one. In the empirical examples we include $p=1$ lags of the endogenous variables.\footnote{We have also experimented with higher lag orders and also found some evidence of signals at lag two for the dataset at hand; see Figures~\ref{fig:ll2} to \ref{fig:ll5b} in the Appendix for an illustration. However, out-of-sample predictive studies favored one lag only (cf.\ Section~\ref{sec:forecast}).} The hyperparameters are chosen as follows: $M_\mu = 10$, $a_0=20$, $b_0=1.5$, $\xi = 1$, $a_\lambda = 0.1$, $c_\lambda = d_\lambda = 1$.

\subsection{Some empirical key features of the model} \label{sec:results}
To provide some intuition on how our modeling approach works in practice, we first estimate a simple one factor model (i.e.\ $q=1$) and investigate several features of our empirical model. In the next section we will perform an extensive forecasting exercise and discuss the optimal number of factors in terms of forecasting accuracy. 

We start by inspecting the posterior distribution of $\boldsymbol{\Lambda}$ and assess what variables load heavily on the latent factor. It is worth emphasizing that most quantities\footnote{Hereby we refer to the one-step-ahead forecast error related to a given time series.} associated with real activity (i.e.\ industrial production and its components, GDP growth, employment measures) load heavily on the factor. Moreover, expectation measures, housing markets, equity prices and spreads also load heavily on the joint factor.

\begin{sidewaysfigure}
\includegraphics[width=\textwidth, trim = 0 20 0 0, clip = TRUE]{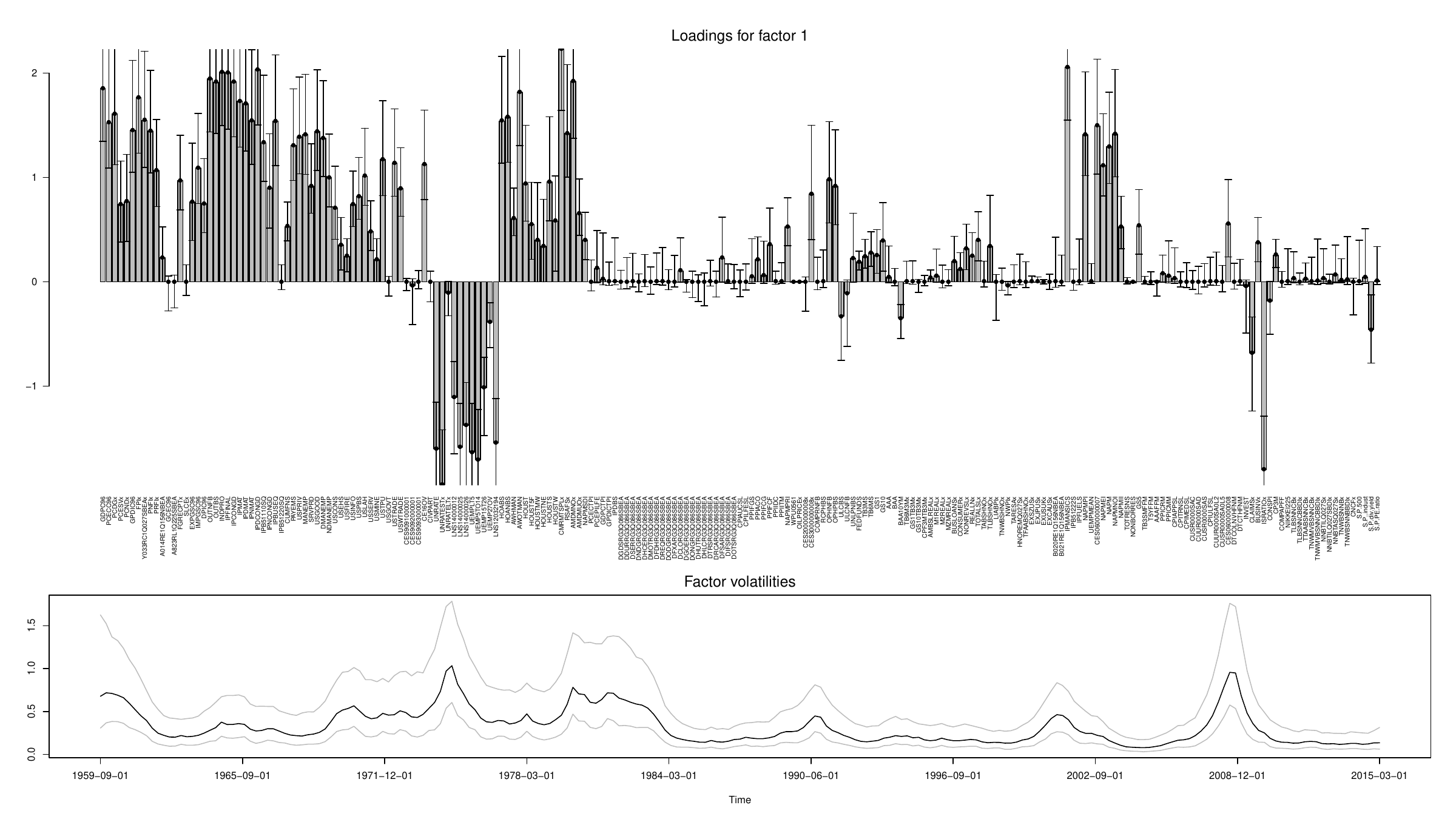}
\caption{$5$th, $50$th, and $95$th posterior percentiles of factor loadings (upper panel) and factor volatility (lower panel).}
\label{fig:vola}
\end{sidewaysfigure}

 To assess whether spikes in the volatility associated with the factor coincide with major economic events, the bottom panel of Figure~\ref{fig:vola} depicts the evolution of the posterior distribution of factor volatility over time. A few findings are worth mentioning. First, volatility spikes sharply during the midst of the 1970s, a period characterized by the first oil price shock and the bankruptcy of Franklin National Bank in 1974. After declining markedly during the second half of the 1970s, the shift in US monetary policy towards aggressively fighting inflation and the second oil price shock again translate into higher macroeconomic uncertainty. Note that from the mid 1980s onward, we observe a general decline in macroeconomic volatility that lasts until the beginning of the 1990s. There we observe a slight increase in volatility possibly caused by the events surrounding the first gulf war. The remaining years up to the beginning of the 2000s has been relatively unspectacular, with volatility levels being muted most of the time. In 2000/2001, volatility again increases due to the burst of the dot-com bubble and the 9/11 terrorist attacks. Finally, we observe marked spikes in volatility during recessionary episodes like the recent financial crisis in 2008. 

\begin{figure}[p]
\includegraphics[width=\textwidth, page=1, trim = 0 0 0 40, clip = TRUE]{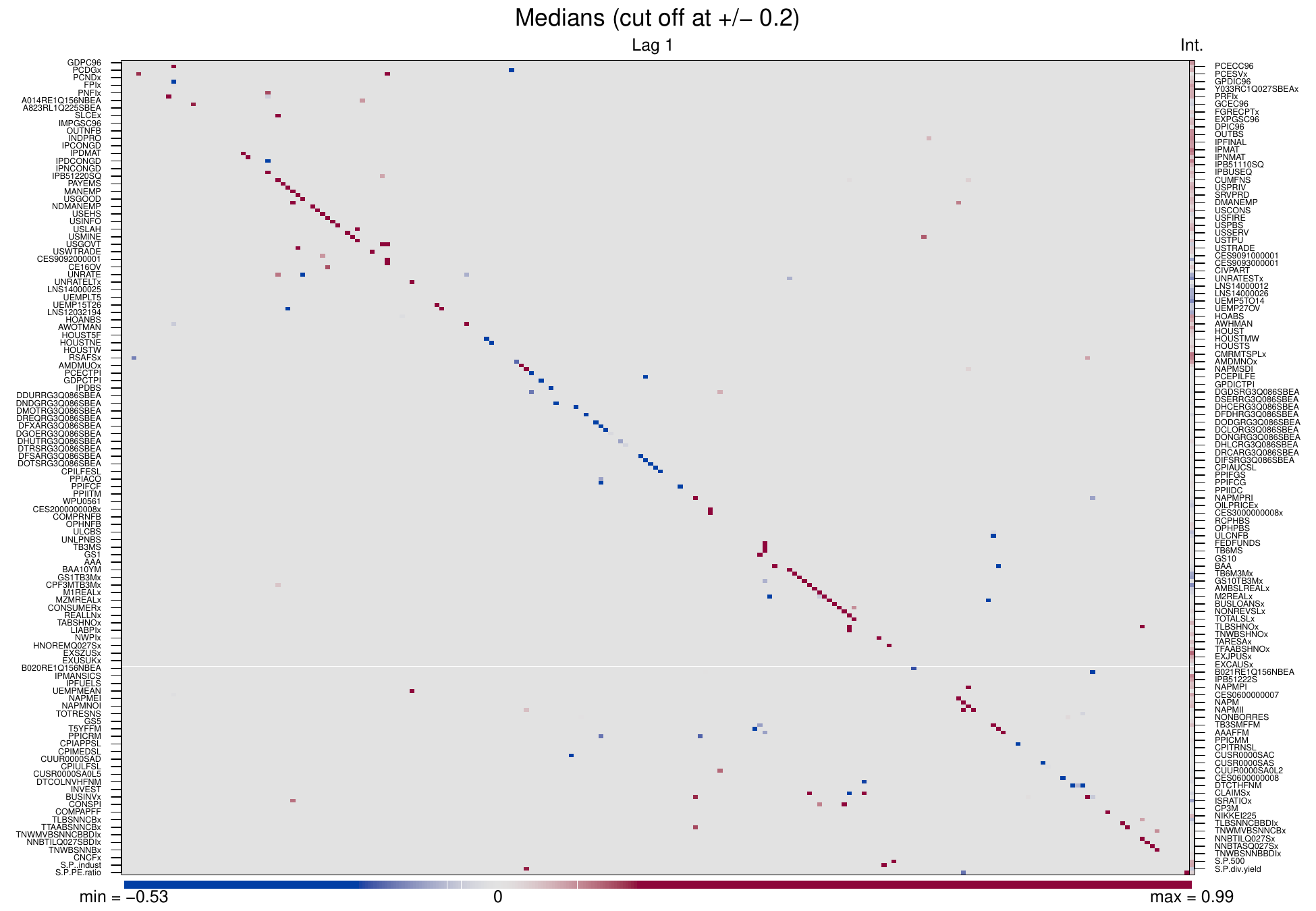}\\
\includegraphics[width=\textwidth, page=1, trim = 0 0 0 40, clip = TRUE]{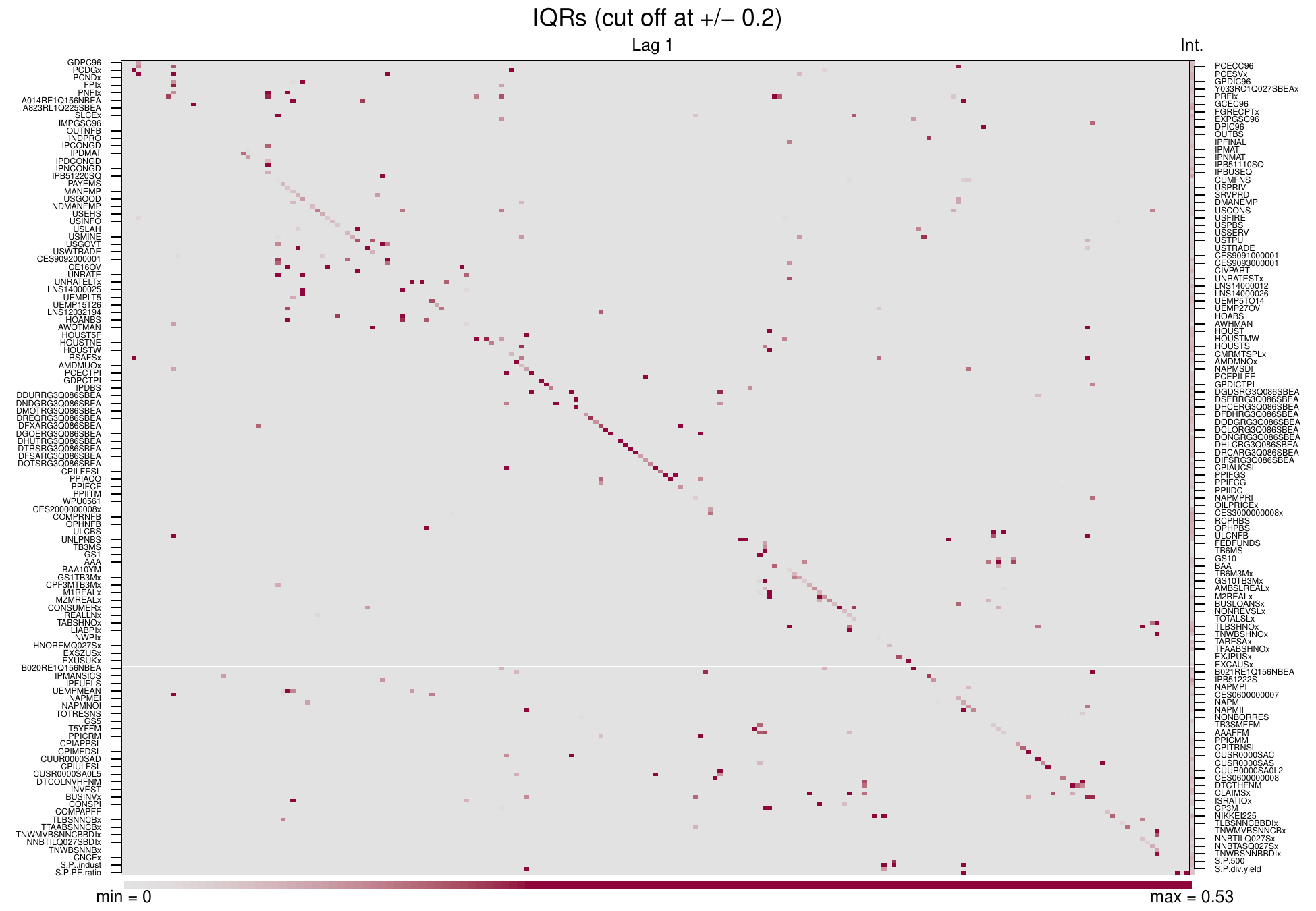}
\caption{Posterior medians (top) and posterior interquartile ranges (bottom) of VAR coefficients, $a = 1/k = 1/216$.}
\label{fig:AK}
\end{figure}

Finally, we assess how well the DL prior with $a=1/k$ performs in shrinking the coefficients in $\boldsymbol{B}$ to zero. The top panel of Figure~\ref{fig:AK} depicts a heatmap that gives a rough feeling on the size of each regression coefficient based on the posterior median of $\boldsymbol{B}$. The bottom panel of Figure~\ref{fig:AK} depicts the posterior interquartile range, providing some evidence on posterior uncertainty.\footnote{Since the corresponding posterior distribution is quite heavy-tailed, using posterior standard deviations, while providing a qualitatively similar picture, tend to be slightly exaggerated.} The DL prior apparently succeeds in shrinking the vast majority of the approximately $50\,000$ coefficients towards zero. Even though not discussed in detail to conserve space, we note that at higher lag orders this very strong shrinkage effect is even more pronounced; see also Figures~\ref{fig:ll2} to \ref{fig:ll5b} in the Appendix. 

\begin{figure}[p]
\includegraphics[width=\textwidth, page=1, trim = 0 0 0 40, clip = TRUE]{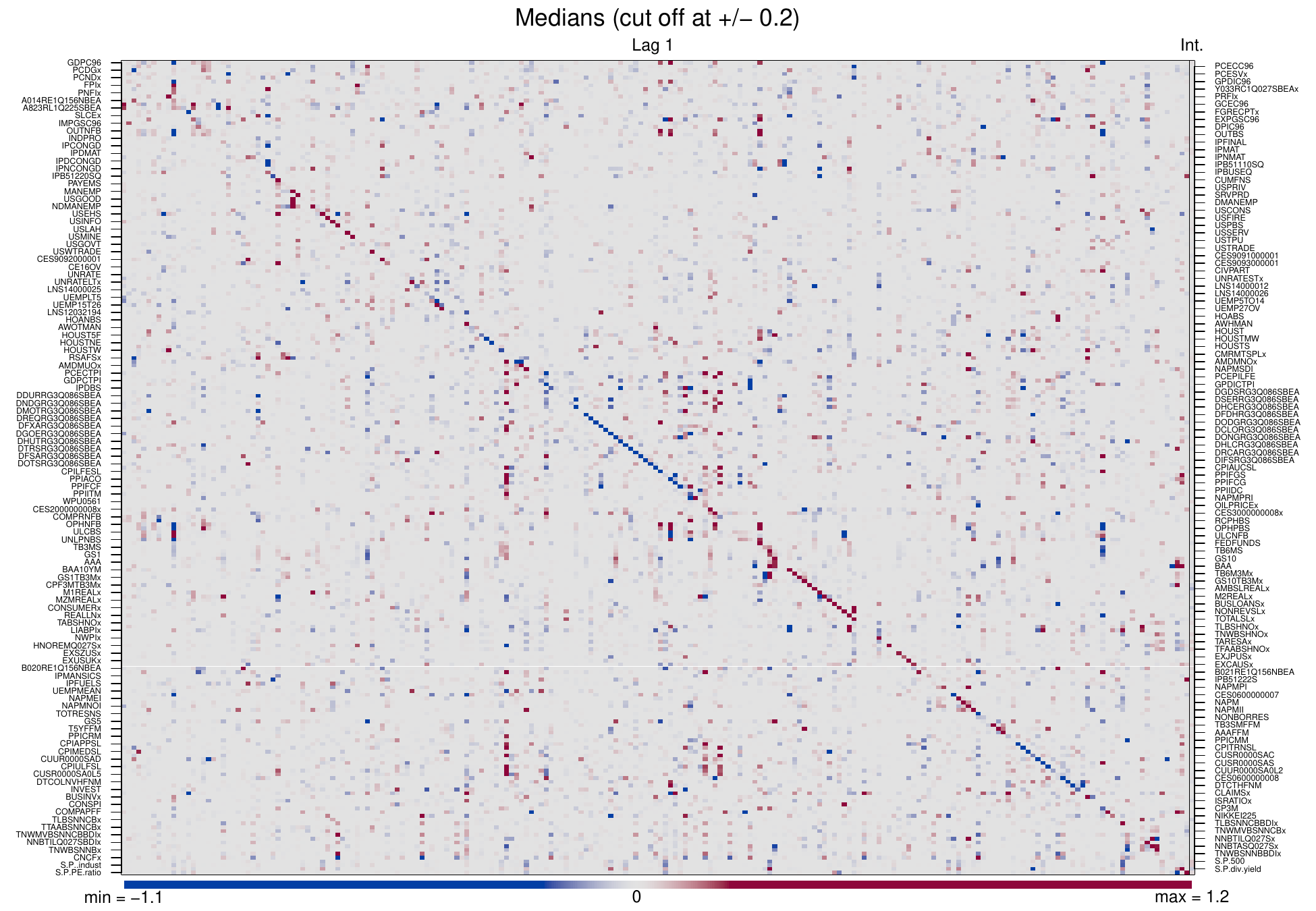}\\
\includegraphics[width=\textwidth, page=1, trim = 0 0 0 40, clip = TRUE]{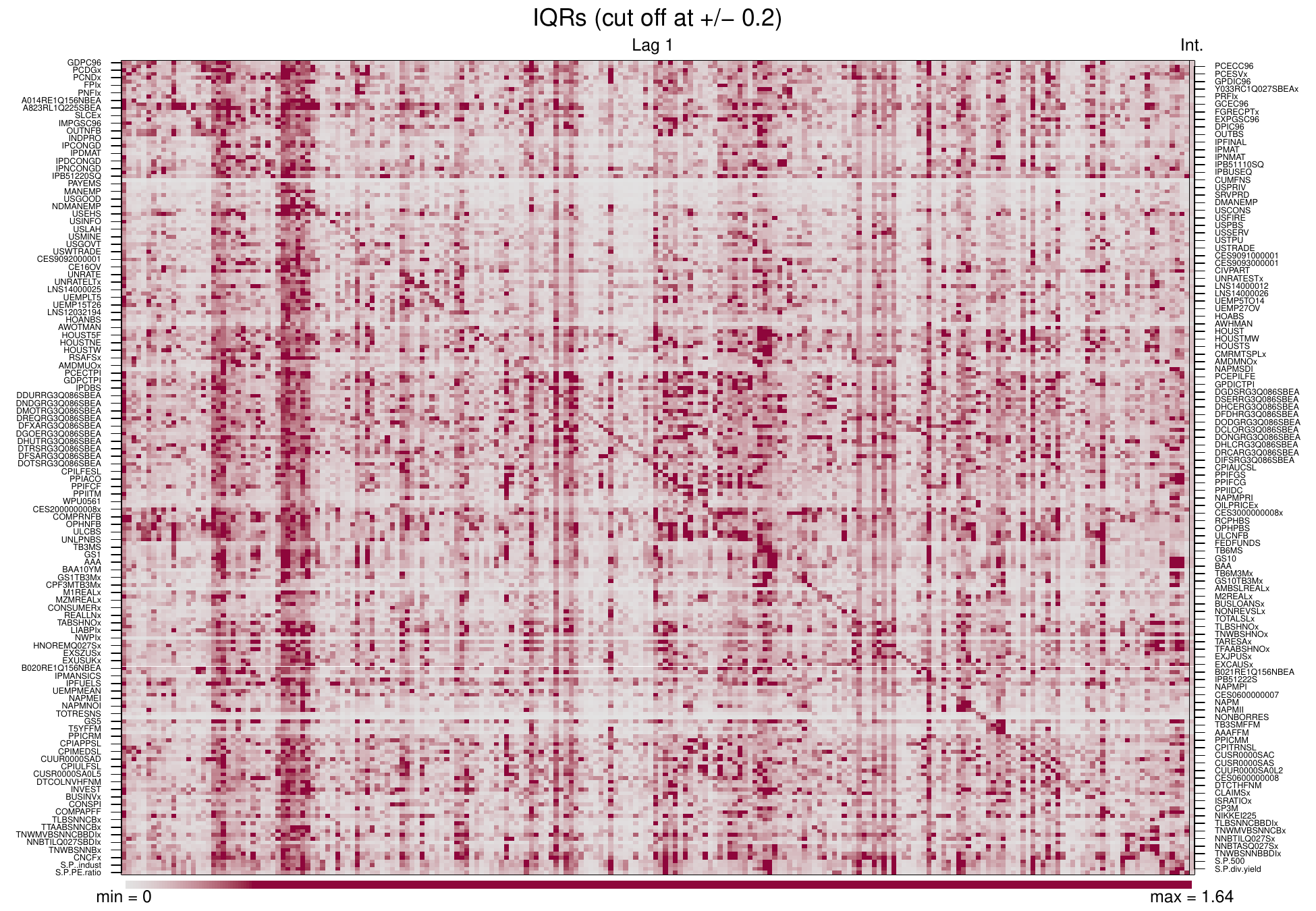}
\caption{Posterior medians (top) and posterior interquartile ranges (bottom) of VAR coefficients, $a = 1/2$.}
\label{fig:A2}
\end{figure}

The top panel of Figure~\ref{fig:A2} displays the posterior median estimates
%for the real data example discussed in Section~\ref{sec:appli} of the main paper
when the shrinkage parameter $a$ is chosen to be $1/2$ \citep[cf.][for a discussion of this choice]{bhattacharya2015dirichlet}. While $a = 1/2$ appears to provide a fair amount of shrinkage in other applications, for our huge dimensional example this prior exerts only relatively little shrinkage and tends to lead to overfitting.
%The diagonal pattern in the first two lags known from Figure~\ref{fig:AK} in the main paper
The diagonal pattern in the first lag appears here as well, but there is a considerable amount of nonzero medians elsewhere. Correspondingly, the interquartile ranges visualized in the bottom panel of Figure~\ref{fig:A2} are also very large compared to those obtained with $a = 1/k$.

Interestingly, for selected time series measuring inflation (both consumer and producer price inflation) we find that lags of monetary aggregates are allowed to load on the respective inflation series. This result points towards a big advantage of our proposed prior relative to standard VAR priors in the Minnesota tradition: while these priors have been shown  to work relatively well in huge dimensions \citep[see][]{banbura2010}, they also display a tendency to overshrink when the overall tightness of the prior is integrated out in a Bayesian framework, effectively pushing the posterior distribution of $\boldsymbol{B}$ towards the prior mean  and thus ruling out patterns observed under the DL prior.

Inspection of the interquartile range also indicates that the proposed shrinkage prior succeeds in reducing posterior uncertainty markedly. Note that the pattern found for the posterior median of $\boldsymbol{B}$ can also be found in terms of the posterior dispersion. We again observe that the coefficients associated with the first, own lag of a given variable are allowed to be non-zero whereas in most other cases the associated posterior is strongly concentrated around zero.%For comparison, we provide heatmaps of posterior medians and interquartile ranges for $a=1/2$ in the Appendix, see Figure~\ref{fig:A2}.

\subsection{Predictive evidence}\label{sec:forecast}
We focus on forecasting  gross domestic product (GDPC96), industrial production (INPRO), total nonfarm payroll (PAYEMS), civilian unemployment rate (UNRATE), new privately owned housing units started (HOUST), consumer price index inflation (CPIAUCSL), producer price index for finished goods inflation (PPIFGS), effective federal funds rate (FEDFUNDS), 10-year treasury constant maturity rate (GS10), U.S./U.K.\ exchange rate (EXUSUKx), and the S\&P 500 (S.P.500). This choice includes the variables investigates by \cite{koop2019bayesian} and some additional important macroeconomic indicators that are commonly monitored by practitioners, resulting in a total of eleven series.

To assess the forecasting performance of our model, we conduct a pseudo out-of-sample forecasting exercise with initial estimation sample ranging from 1959:Q3 to 1990:Q2. Based on this estimation period, we compute one-quarter-ahead predictive densities for the first period in the hold-out (i.e.\ 1990:Q3). After obtaining the corresponding predictive densities and evaluating the corresponding log predictive likelihoods, we expand the estimation period and re-estimate the model. This procedure is repeated $100$ times until the final point of the full sample is reached. The quarterly scores obtained this way are  then accumulated.

Our model with  $q \in \{0,1\dots,4\}$ factors is benchmarked against the prior model, a pure factor stochastic volatility (FSV) model with conditional mean equal to zero (i.e. $\boldsymbol{B}=\boldsymbol{0}_{m \times k})$. In what follows we label this specification  FSV $0$. To assess the merits of the proposed shrinkage prior vis-\'{a}-vis a Minnesota prior and a NG shrinkage prior we also include the models described in Section \ref{sec:simstudy}. Moreover, we include  two models that impose the restriction that $\boldsymbol{A}_1=\boldsymbol{I}_m$ and $\boldsymbol{A}_1=0.8 \times \boldsymbol{I}_m$ while $\boldsymbol{A}_j$ for $j>1$ are set equal to zero matrices in both cases. The first model, labeled FSV $1$, assumes that the  conditional mean of $\boldsymbol{y}_t$ follows a random walk process and the second specification, denoted as FSV $0.8$, imposes the restriction that the variables in $\boldsymbol{y}_t$ feature a rather strong degree of persistence but are stationary.  The exercise serves to evaluate whether it pays off to impose a VAR structure on the first moment of the joint density of our data and to assess how many factors are needed to obtain precise multivariate density predictions for our eleven variables of interest.

\begin{table}[t]
\centering
\caption{\label{tab:relCumLPSsome}Average log predictive scores for the number of factors $q \in \{0, 1, \dots, 5\}$ in various VAR-FSV specifications as well as pure FSV models. Estimation and prediction is conducted on all $m=215$ component series; the predictive density is then evaluated on the set of $11$ variables of interest. Larger numbers indicate better joint predictive density performance.} 
\begin{tabular}{rrrrrrr}
  \hline
 & 0 & 1 & 2 & 3 & 4 & 5 \\ 
  \hline
VAR($1$)-FSV DL($1/2$) & -14.79 & -13.45 & -12.74 & -12.53 & -11.85 & -11.44 \\ 
  VAR($1$)-FSV DL($1/k$) & -10.51 & -9.75 & -9.15 & -9.15 & -9.17 & -9.43 \\ 
  VAR($1$)-FSV DL($1/K$) & -10.54 & -9.63 & -9.22 & -9.26 & -9.22 & -9.38 \\ 
  VAR($1$)-FSV NG($1$) & -10.76 & -10.26 & -9.72 & -9.60 & -9.68 & -9.86\\ 
  VAR($1$)-FSV NG($0.1$) & -10.55 & -9.95 & -9.31 & -9.26 & -9.47 & -9.46 \\ 
  VAR($1$)-FSV Min($0.01$) & -10.99 & -10.37 & -9.89 & -9.88 & -9.98 & -10.24 \\ 
  VAR($1$)-FSV Min($0.001$) & -12.15 & -11.21 & -11.08 & -10.62 & -10.56 & -10.83 \\ 
  VAR($2$)-FSV DL($1/k$) & -10.50 & -9.88 & -9.22 & -9.38 & -9.28 & -9.43 \\ 
  VAR($2$)-FSV NG($0.1$) & -10.47 & -10.04 & -9.44 & -9.29 & -9.40 & -9.49 \\ 
  VAR($5$)-FSV DL($1/k$) & -10.70 & -9.97 & -9.78 & -9.21 & -9.44 & -9.53 \\ 
  VAR($5$)-FSV NG($0.1$) & -10.61 & -10.01 & -9.64 & -9.51 & -9.50 & -9.52 \\ 
  FSV 0 & -12.08 & -11.08 & -11.02 & -10.66 & -10.61 & -10.71 \\ 
  FSV 0.8 & -11.78 & -11.32 & -11.06 & -10.87 & -10.96 & -11.25 \\ 
  FSV 1 & -11.71 & -11.28 & -10.95 & -10.95 & -11.07 & -11.18 \\ 
   \hline
\end{tabular}
\end{table}

Overall log predictive scores (LPSs) are summarized in Table~\ref{tab:relCumLPSsome}. An immediate finding is that ignoring the error covariance structure (using zero factors) produces rather inaccurate forecasts for all models considered. While a single factor model improves predictive accuracy by a large margin, allowing for more factors (i.e.\ even more flexible modeling of the covariance structure) further increases the forecasting performance. For this specific exercise, we identify two or three factors to be a reasonable choice for most models when the joint log predictive scores of the aforementioned variables are considered. We would like to stress that this choice critically depends on the number of variables we include in our prediction set. If we focus attention on the marginal predictive densities (i.e.\ the univariate predictive densities obtained after integrating out the remaining elements in $\boldsymbol{y}_t$) we find that fewer or even no factors receive more support (see Table~\ref{tab:univLPS}), whereas in the case of higher dimensional prediction sets more than two factors lead to more accurate density predictions \citep[cf.][for an investigation of this issue in the context of a standard FSV model]{kas-etal:eff}. As a general remark, we note that identifying the optimal number of factors in high-dimensional FSV models is a challenging problem in practice. Using the deviance information criterion \citep[DIC, cf.][]{chan2016fast} may be an option but is likely to be unstable in very high dimensions. The approach adopted in the paper at hand, namely the decomposition of the marginal likelihood into predictive likelihoods \citep[cf.][]{geweke2010comparing} tends to be more stable, in particular when interest ist placed on predicting subsets only. Moreover, it can be trivially parallelized, thus becoming computationally feasible on high performance computing infrastructures.

Considering forecasting accuracy across models reveals that our proposed VAR($1$)-FSV with a DL($1/k$) prior displays excellent forecasting capabilities, outperforming all competitors. Amongst the VAR($1$) models, DL($1/K$) and NG($0.1$) also do well, and the Bayesian Lasso (NG($1$)) as well as the Minnesota prior with medium shrinkage (Min($0.01$)) show decent performance. Clearly, DL($1/2$) overfits and Min($0.001$) overshrinks. Note that higher lag orders seem to rarely increase predictive accuracy.
However, comparing the differences between the benchmark pure FSV models and the VAR-FSV models considered, we find that explicitly modeling the conditional mean improves the forecasting accuracy in practically all cases.
%Furthermore, we conjecture that the inclusion of the factors has pronounced effects on the estimates of $\boldsymbol{B}$ and this could potentially lead to an increase in predictive accuracy. Intuitively, this might be explained by the fact that if the underlying data generating process suggests that  $\boldsymbol{B}$ is time-varying, inclusion of the factors might alleviate issues associated with model misspecification by, at least to a certain extent, controlling for structural breaks in $\boldsymbol{B}$.
%Looking at the differences between both specifications that assume that the conditional mean of $\boldsymbol{y}_t$ is highly persistent and the VAR-FSV indicates that after including at least a single factor, the VAR-FSV and the FSV $0$ outperform both, the FSV $0.8$ and the FSV $1$, competing specifications. 

%For this prediction exercise, four factors appear to be a reasonable choice.\footnote{Note that the optimal number of factors depends on the numbers of variables to be (jointly) forecast. Without presenting the results in detail due to space restrictions, considering univariate predictive densities tends to favor fewer or even no factors, higher dimensional prediction sets support more factors.}

\begin{figure}[t]
\centering
\includegraphics[width = \textwidth]{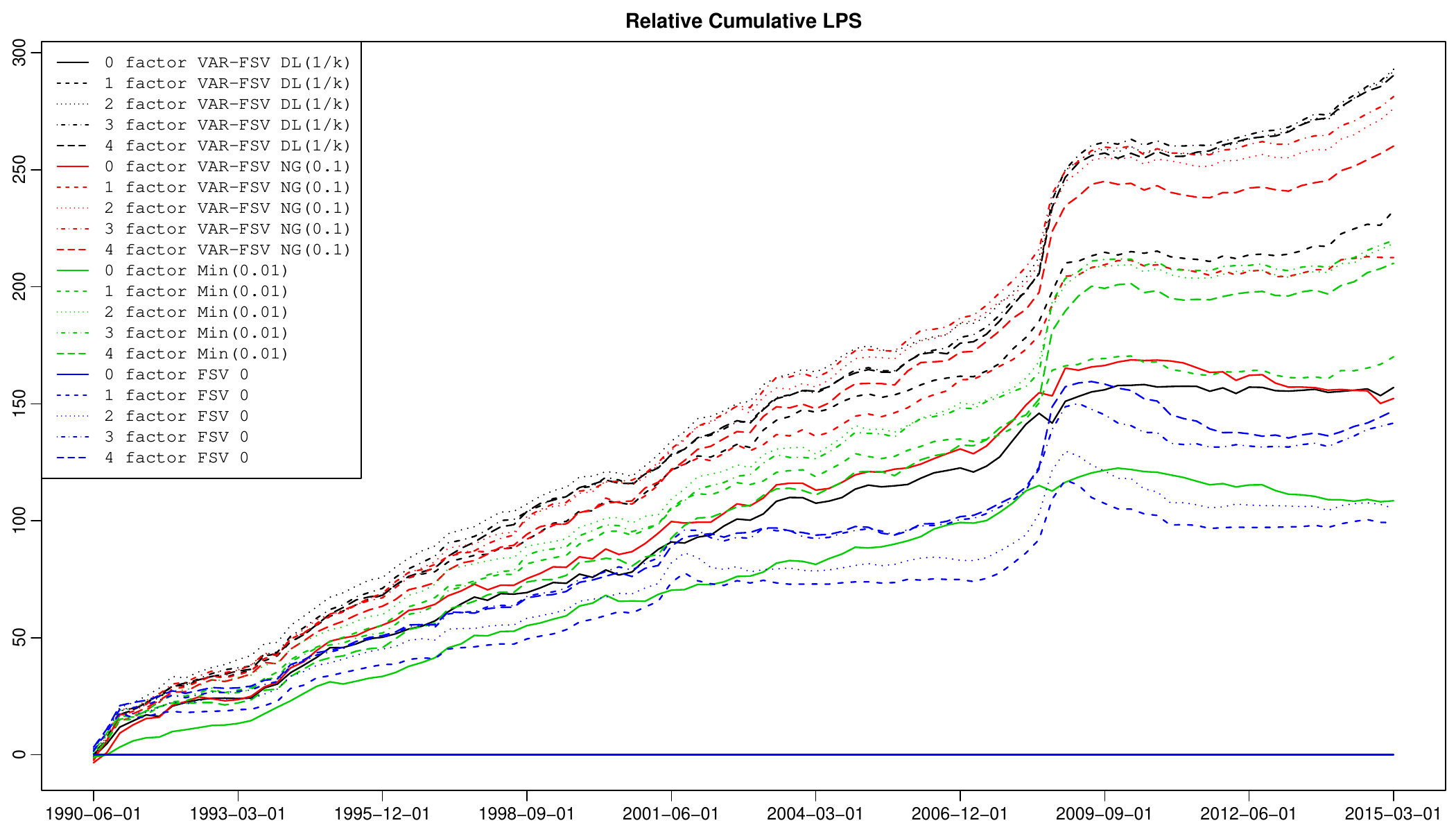}
\caption{Cumulative log predictive scores, relative to a zero-mean model with independent stochastic volatility components for all component series. Higher values correspond to better one-quarter-ahead density predictions up to the corresponding point in time.}
\label{fig:relCumLPSsome}
\end{figure}

To investigate whether forecasting performance is homogeneous over time,  Figure~\ref{fig:relCumLPSsome} visualizes the cumulative LPSs relative to the zero-factor FSV model over time.
%To keep the presentation feasible, we only include the three strongest competitors in the graphical display but note that the pattern of the LPS appears to be similar across specifications.
%Since the FSV $0$ model proves to be the strongest competitor and the pattern of the LPS appears to be similar across specifications, we exclude the FSV $0.8$ and the FSV $1$ from this figure.
The benefit of the flexible SV structure in the VAR residuals is particularly pronounced during the 2008 financial crisis which can be seen by comparing the solid lines to the broken lines. During this period, time-varying covariance modeling appears to be of great importance and the performance of models that ignore contemporaneous dependence deteriorates. This finding is in line with \cite{kas:spa} who reports analogous results for US asset returns. The increase in predictive accuracy can be traced back to the fact that within an economic downturn, the correlation structure of our dataset changes markedly, with most indicators that measure real activity sharply declining in lockstep. A model that takes contemporaneous cross-variable linkages seriously is thus able to fully exploit such behavior which in turn improves predictions. 

\begin{table}[t]
\caption{\label{tab:}\label{tab:univLPS} Average univariate log predictive scores for inflation (CPIAUCSL), short-term interest rates (FEDFUNDS), and output growth (GDPC96) with $q \in \{0,1,2\}$ factors.}

\centering
\begin{adjustbox}{max width=\textwidth}
\begin{tabular}{lrrrrrrrrr}
\toprule
\multicolumn{1}{c}{ } & \multicolumn{3}{c}{CPIAUCSL} & \multicolumn{3}{c}{FEDFUNDS} & \multicolumn{3}{c}{GDPC96} \\
\cmidrule(l{3pt}r{3pt}){2-4} \cmidrule(l{3pt}r{3pt}){5-7} \cmidrule(l{3pt}r{3pt}){8-10}
  & 0 & 1 & 2 & 0 & 1 & 2 & 0 & 1 & 2\\
\midrule
%VAR($1$)-FSV DL($1/2$) & -1.23 & -1.14 & -1.08 & -1.42 & -1.39 & -1.36 & -0.71 & -0.71 & -0.70\\
VAR($1$)-FSV DL($1/k$) & -1.03 & -1.11 & -1.13 & -1.26 & -1.26 & -1.24 & 0.08 & 0.05 & 0.03\\
%VAR($1$)-FSV DL($1/K$) & -1.05 & -1.12 & -1.15 & -1.29 & -1.24 & -1.22 & 0.07 & 0.04 & 0.02\\
%VAR($1$)-FSV NG($1$) & -1.00 & -1.09 & -1.12 & -1.28 & -1.26 & -1.26 & -0.23 & -0.25 & -0.25\\
VAR($1$)-FSV NG($0.1$) & -1.00 & -1.07 & -1.10 & -1.28 & -1.26 & -1.22 & -0.10 & -0.12 & -0.15\\
%VAR($1$)-FSV Min($0.01$) & -1.01 & -1.17 & -1.17 & -1.27 & -1.26 & -1.26 & -0.25 & -0.35 & -0.37\\
%VAR($1$)-FSV Min($0.001$) & -1.12 & -1.19 & -1.21 & -1.28 & -1.28 & -1.27 & -0.20 & -0.40 & -0.45\\
VAR($2$)-FSV DL($1/k$) & -1.02 & -1.12 & -1.14 & -1.25 & -1.27 & -1.22 & 0.04 & -0.01 & -0.01\\
VAR($2$)-FSV NG($0.1$) & -1.00 & -1.11 & -1.12 & -1.26 & -1.23 & -1.25 & -0.14 & -0.16 & -0.14\\
VAR($5$)-FSV DL($1/k$) & -1.05 & -1.16 & -1.16 & -1.29 & -1.26 & -1.26 & -0.02 & -0.10 & -0.13\\
VAR($5$)-FSV NG($0.1$) & -1.00 & -1.09 & -1.13 & -1.29 & -1.27 & -1.25 & -0.19 & -0.18 & -0.21\\
\bottomrule
\end{tabular}
\end{adjustbox}
\end{table}

Up to this point, we focused exclusively on the joint performance of our model for the specific set of variables considered. To gain a deeper understanding on how our model performs for relevant selected quantities, Table~\ref{tab:univLPS} displays  marginal LPSs for the two most promising prior specifications with one, two, and five lags. The variables we consider are inflation (CPIAUCSL), short-term interest rates (FEDFUNDS), and output growth (GDPC96).

In contrast to the findings based on joint LPSs, we observe that models without a factor structure tend to perform better than models that set $q>0$, with the exception of interest rates where all models predict more or less equally badly. This finding corroborates our conjecture stated above, implying that if the set of focus variables is subsequently enlarged, more factors are necessary in order to obtain precise density predictions. Here, we only focus on marginal model performance, implying that for each variable, contemporaneous relations between the elements in $\bm{y}_t$ are integrated out. This, in turn, implies that the additional gain in model flexibility is offset by the comparatively larger number of parameters.
Concerning the difference between VAR priors, it appears that NG slightly outperforms DL for inflation whereas DL is superior when it comes to predicting output growth.
%This finding holds true for inflation and output growth. For short-term interest rates, a different pattern emerges. Considering the marginal LPS for FEDFUNDS across different $q$ suggests almost no differences between choosing $q=0$ and $q=2$.

%For brevity, and in light of the rather weak forecasting performance of the models that do not explicitly model the conditional mean, we focus exclusively on the different VAR models considered.

%A careful analysis of whether the VAR structure on the mean actually improves forecasting performance leads to an affirmative conclusion only after the turn of the millennium, where models featuring the vector autoregressive mean structure tend to do better than their zero-mean counterparts. Note that the period surrounding the burst of the dot-com bubble and the 9/11 terrorist attacks appears to be the first time frame where VAR models sharply improve upon the FSV specification. After 2002, most models that explicitly cater for the conditional mean continue to improve upon the different FSV specifications. The accuracy gains are particularly pronounced  through the 2008 financial crisis, with the gap widening even further. This trend continues until the end of the sample if the VAR residuals are allowed to have two or more latent factors.

\subsection{A note on the computational burden}\label{sec:comptime}
\begin{figure}[t]
\centering
\includegraphics[width = .9\textwidth]{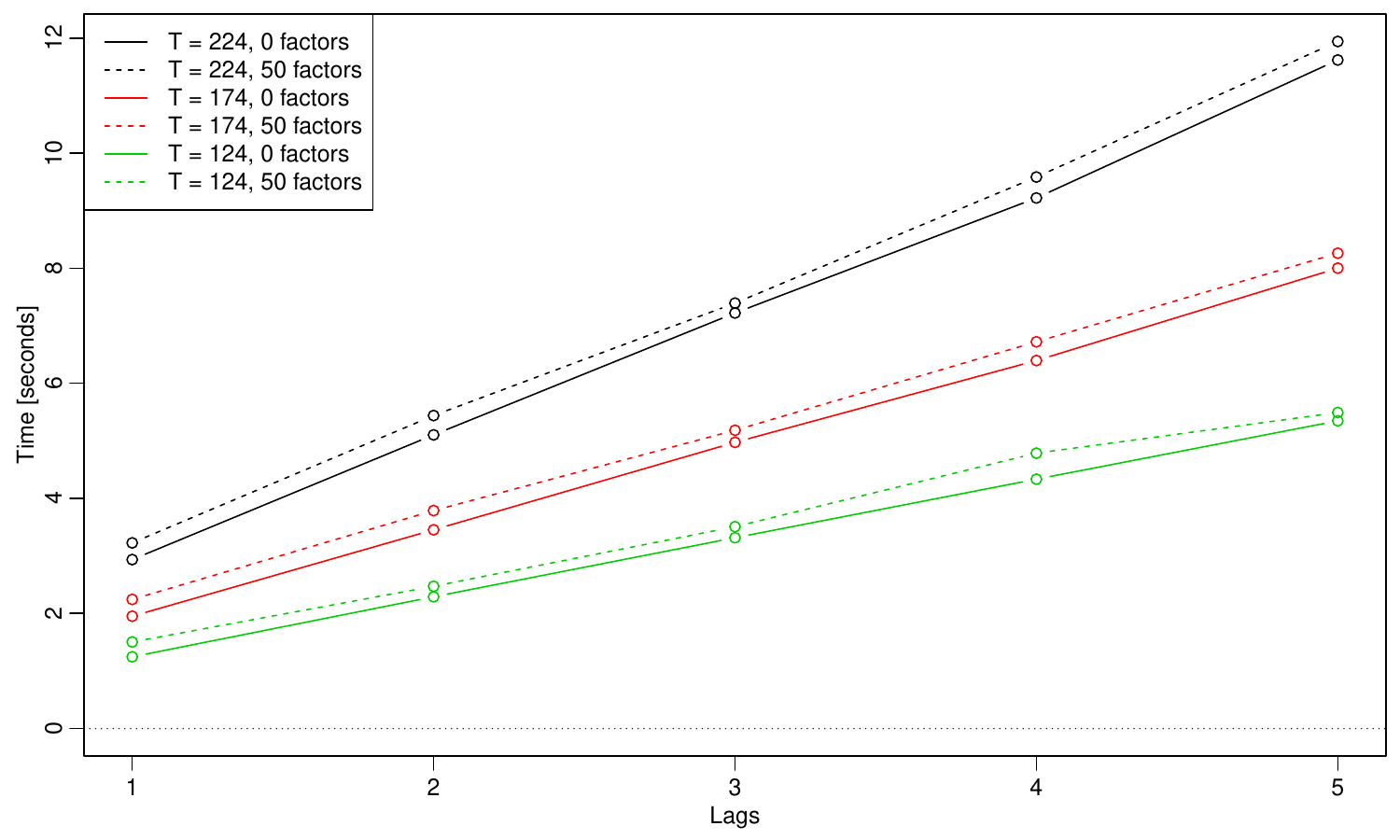}
\caption{Empirical CPU times for each MCMC iteration on a standard laptop computer using one core. Time series lengths: $T \in \{124, 174, 224\}$; numbers of latent factors: $q \in \{0, 50\}$.}\label{fig:timings}
\end{figure}

Even though the efficient sampling schemes outlined in this paper help to overcome absolutely prohibitive computational burdens, the CPU time needed to perform fully Bayesian inference in a model of this size can still be considered substantial. In what follows we shed light on the estimation time required and how it is related to the length of the time series $T$, the lag length $p$ and to the number of latent factors $q \in \{0, 50\}$. Figure~\ref{fig:timings} shows the time needed to perform a single draw from the joint posterior distribution of the $215 + 215^2p$ coefficients and their corresponding $2(215 + 215^2p) + 1$ auxiliary shrinkage quantities, the $qT$ factor realizations and the associated $215q$ loadings, alongside $(T+1)(215 + q)$ latent volatilities with their corresponding $645 + 2q$ parameters. This amounts to $166\,841$ random draws for the smallest model considered (one lag, no factors, $T=124$) and $776\,341$ random draws for the largest model ($5$ lags, $50$ factors, $T=224$) at each MCMC iteration.

As mentioned above, the computation time rises approximately linearly with the number of lags included. Dotted lines indicate the time in seconds needed to perform a single draw from a model with $50$ factors included while solid lines refer to the time needed to estimate a model without factors and a diagonal time-varying variance-covariance matrix $\boldsymbol{\Sigma}_t$. Interestingly, the additional complexity when moving from a model without factors to a highly parameterized model with $50$ factors appears to be negligible, increasing the time needed by a fraction of a second on average. The important role of the length of the sample can be seen by comparing the green, red and black lines. The time necessary to perform a simple MCMC draw quickly rises with the length of our sample, consistent with the statements made in Section~\ref{sec:comp}. This feature of our algorithm, however, is convenient especially when researchers are interested in combining many short time series or performing recursive forecasting based on a tiny initial estimation sample.

\section{Closing remarks}\label{sec:concl}
In this paper we propose an alternative route to estimate huge dimensional VAR models that allow for time-variation in the error variances. The Dirichlet-Laplace prior, a recent variant of a global-local shrinkage prior, enables us to heavily shrink the parameter space towards the prior model while providing enough flexibility that individual regression coefficients are allowed to be unrestricted. This prior setup alleviates overfitting issues generally associated with large VAR models. To cope with computational issues we assume that the one-step-ahead forecast errors of the VAR feature a factor stochastic volatility structure that enables us to perform equation-by-equation estimation, conditional on the loadings and the factors. Since posterior simulation of each equation's autoregressive parameters involves manipulating large matrices, we implement an alternative recent algorithm that improves upon existing methods by large margins, rendering a fully fledged Bayesian estimation of truly huge systems possible.

In an empirical application we first present various key features of our approach based on a single factor model. This single factor which summarizes the joint dynamics of the VAR errors can be interpreted as an uncertainty measure that closely tracks observed factors such as the volatility index. The question whether such a simplistic structure proves to be an adequate representation of the time-varying covariance matrix naturally arises and we thus provide a detailed forecasting exercise to evaluate the merits of our approach relative to the prior model and a set of competing  models with a different number of latent factors in the errors. 

Finally, three potential extensions are worth mentioning.
First, given the fact that systematic and in-depth empirical comparisons of the various recently developed roads towards handling high-dimensional VARs with time-varying contemporaneous covariance in a Bayesian framework (VAR-FSV, VAR-Cholesky-SV, compressed VAR-SV, etc.) is still missing and it is not clear whether one of these models turns out to dominate the others for all points in time, one could consider to average/select dynamically.
Second, note that it is trivial to relax the assumption of symmetry for the DL components. In the context of VARs, this might be of particular interest for distinguishing diagonal ($a_D$ large) from off-diagonal ($a_O$ small) elements in the spirit of the Minnesota prior or increasing the amount of shrinkage with increasing lag order \citep[cf.][for a similar setup in the context of the normal-gamma shrinkage prior]{huber2016adaptive}.
Third, we would like to stress that our approach could also be used to estimate huge dimensional time-varying parameter VAR models with stochastic volatility. To cope with the computational difficulties associated with the vast state space, a possible approach could be to rely on an additional layer of hierarchy that imposes a (dynamic) factor structure on the time-varying autoregressive coefficients in the spirit of \cite{eisenstat2018reducing} and thus reduce the computational burden considerably.

\printbibliography

\begin{appendix}
 \section*{Appendix: Further illustrations}
\label{suppA}
\begin{figure}
 \includegraphics[page=1, width = .49\textwidth, trim = 60 0 52 70, clip = TRUE]{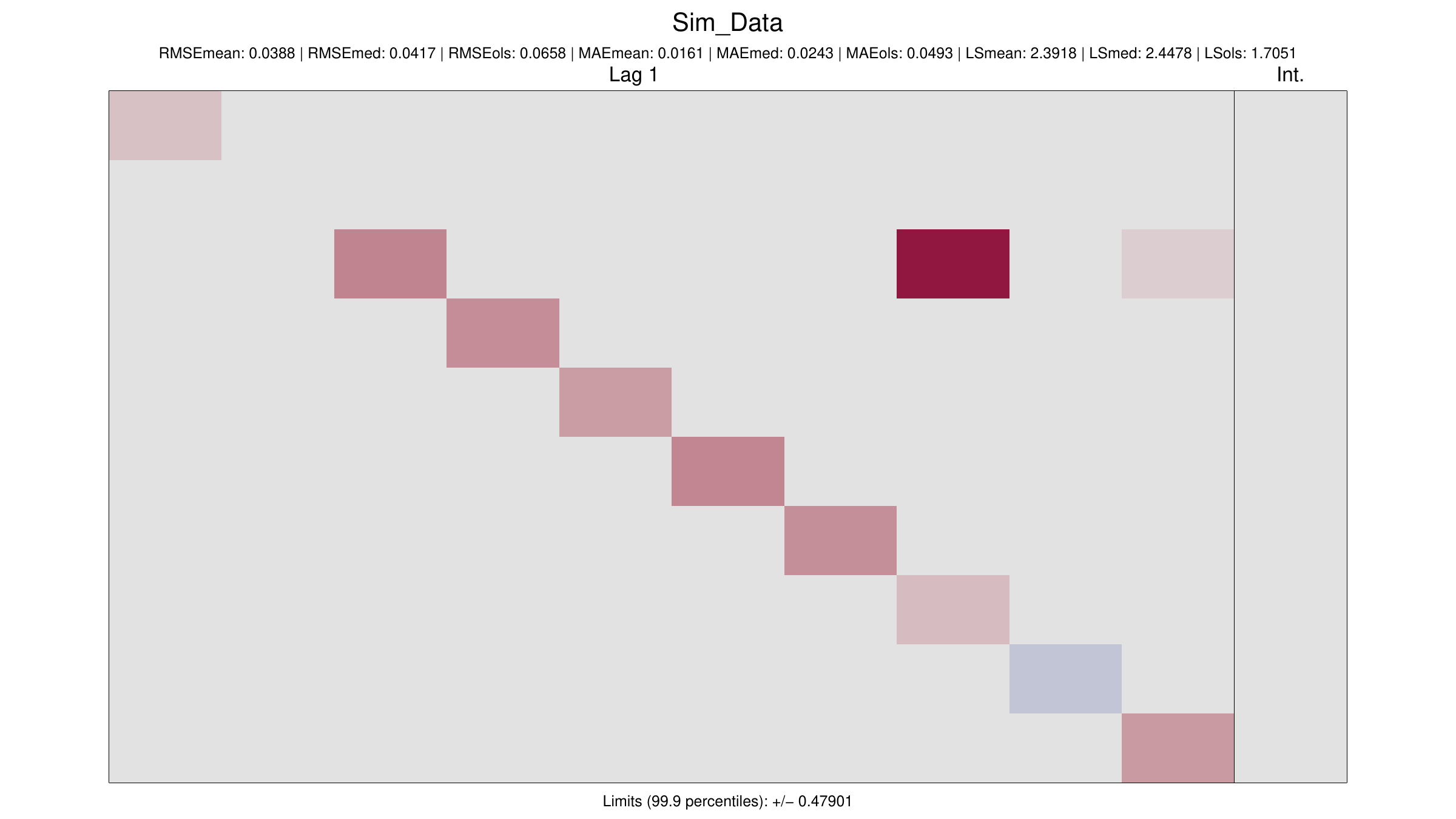}
 \includegraphics[page=4, width = .49\textwidth, trim = 60 0 52 70, clip = TRUE]{sparse_T250_m10/DL2/001_heatmaps.pdf} \\
 \includegraphics[page=2, width = .49\textwidth, trim = 60 0 52 70, clip = TRUE]{sparse_T250_m10/DL2/001_heatmaps.pdf}
 \includegraphics[page=2, width = .49\textwidth, trim = 60 0 52 70, clip = TRUE]{sparse_T250_m10/DL11/001_heatmaps.pdf} \\
 \includegraphics[page=2, width = .49\textwidth, trim = 60 0 52 70, clip = TRUE]{sparse_T250_m10/NG1/001_heatmaps.pdf}
 \includegraphics[page=2, width = .49\textwidth, trim = 60 0 52 70, clip = TRUE]{sparse_T250_m10/NG10/001_heatmaps.pdf} \\
 \includegraphics[page=2, width = .49\textwidth, trim = 60 0 52 70, clip = TRUE]{sparse_T250_m10/M1000/001_heatmaps.pdf}
 \includegraphics[page=2, width = .49\textwidth, trim = 60 0 52 70, clip = TRUE]{sparse_T250_m10/M10000/001_heatmaps.pdf}
 \caption{Exemplary visualization of the true and estimated VAR coefficients in the \emph{sparse} scenario where $T=250$ and $m=10$. Top left: DGP. Top right: OLS estimates. Second row: DL prior with $a_{DL} = 1/2$ (left) and $a_{DL} = 1/k = 1/11$ (right). Third row: NG prior with $a_{NG} = 1$ (left) and $a_{NG} = 1/10$ (right). Fourth row: Minnesota prior with $a_{M} = 1/1000$ (left) and $a_{M} = 1/10000$ (right).}
 \label{simstudyexample1}
\end{figure}

\begin{figure}
 \includegraphics[page=1, width = .49\textwidth, trim = 60 0 52 70, clip = TRUE]{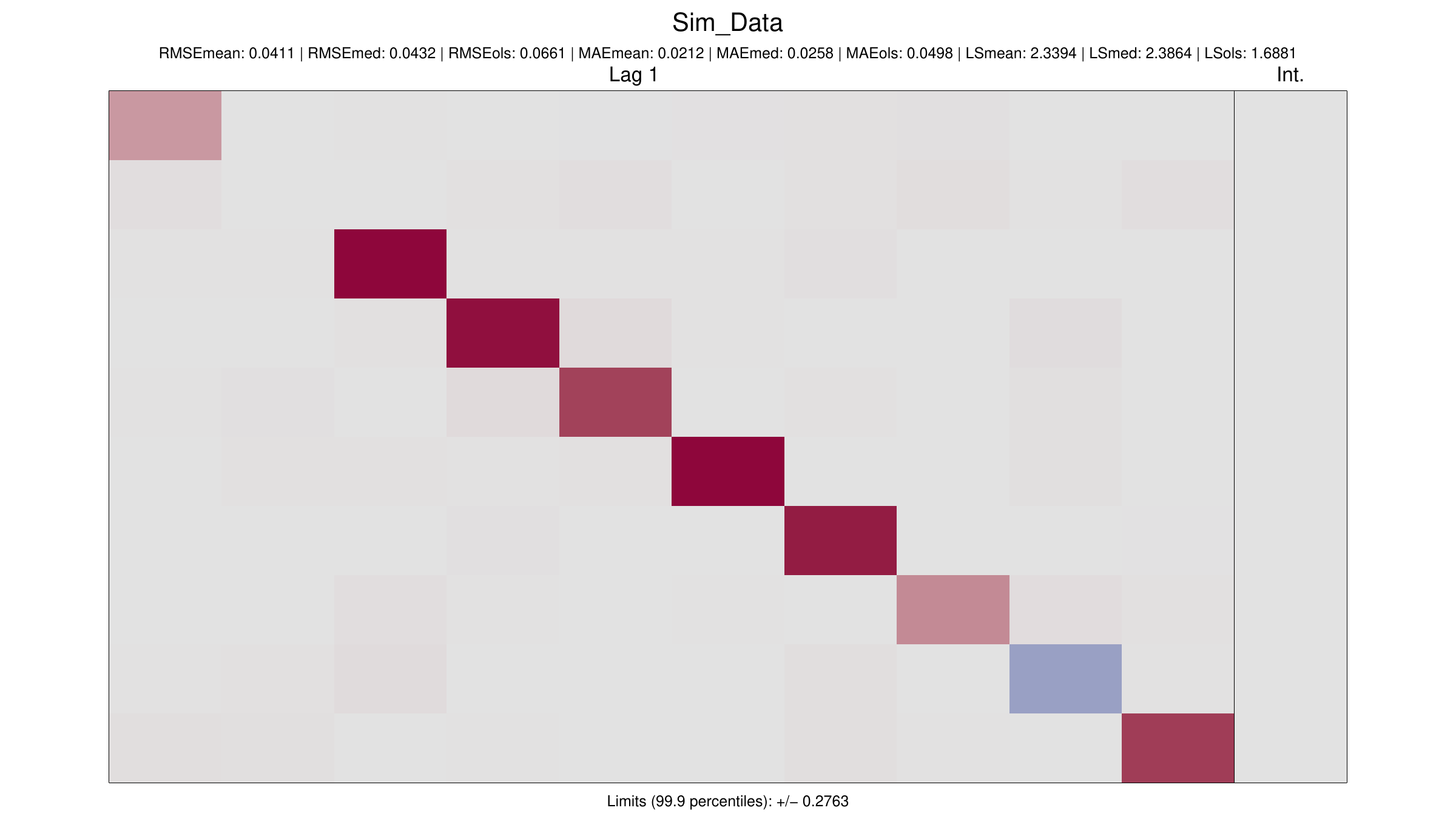}
 \includegraphics[page=4, width = .49\textwidth, trim = 60 0 52 70, clip = TRUE]{dense_T250_m10/DL2/001_heatmaps.pdf} \\
 \includegraphics[page=2, width = .49\textwidth, trim = 60 0 52 70, clip = TRUE]{dense_T250_m10/DL2/001_heatmaps.pdf}
 \includegraphics[page=2, width = .49\textwidth, trim = 60 0 52 70, clip = TRUE]{dense_T250_m10/DL11/001_heatmaps.pdf} \\
 \includegraphics[page=2, width = .49\textwidth, trim = 60 0 52 70, clip = TRUE]{dense_T250_m10/NG1/001_heatmaps.pdf}
 \includegraphics[page=2, width = .49\textwidth, trim = 60 0 52 70, clip = TRUE]{dense_T250_m10/NG10/001_heatmaps.pdf} \\
 \includegraphics[page=2, width = .49\textwidth, trim = 60 0 52 70, clip = TRUE]{dense_T250_m10/M1000/001_heatmaps.pdf}
 \includegraphics[page=2, width = .49\textwidth, trim = 60 0 52 70, clip = TRUE]{dense_T250_m10/M10000/001_heatmaps.pdf}
 \caption{Exemplary visualization of the true and estimated VAR coefficients in the \emph{dense} scenario where $T=250$ and $m=10$. Top left: DGP. Top right: OLS estimates. Second row: DL prior with $a_{DL} = 1/2$ (left) and $a_{DL} = 1/k = 1/11$ (right). Third row: NG prior with $a_{NG} = 1$ (left) and $a_{NG} = 1/10$ (right). Fourth row: Minnesota prior with $a_{M} = 1/1000$ (left) and $a_{M} = 1/10000$ (right).}
 \label{simstudyexample2}
\end{figure}

\begin{figure}
 \includegraphics[page=1, width = .49\textwidth, trim = 60 0 52 70, clip = TRUE]{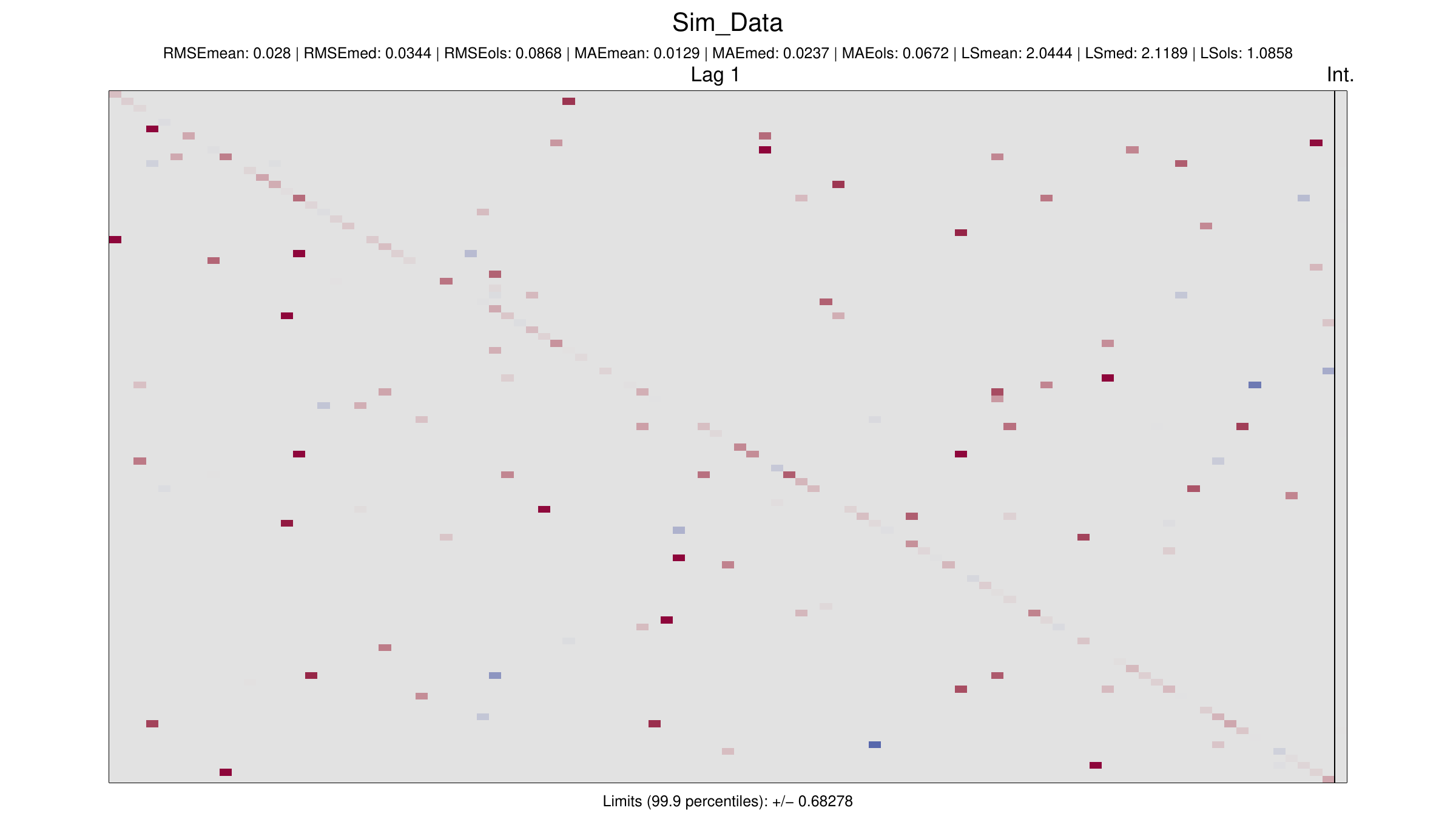}
 \includegraphics[page=4, width = .49\textwidth, trim = 60 0 52 70, clip = TRUE]{sparse_T250_m100/DL2/001_heatmaps.pdf} \\
 \includegraphics[page=2, width = .49\textwidth, trim = 60 0 52 70, clip = TRUE]{sparse_T250_m100/DL2/001_heatmaps.pdf}
 \includegraphics[page=1, width = .49\textwidth, trim = 60 0 52 70, clip = TRUE]{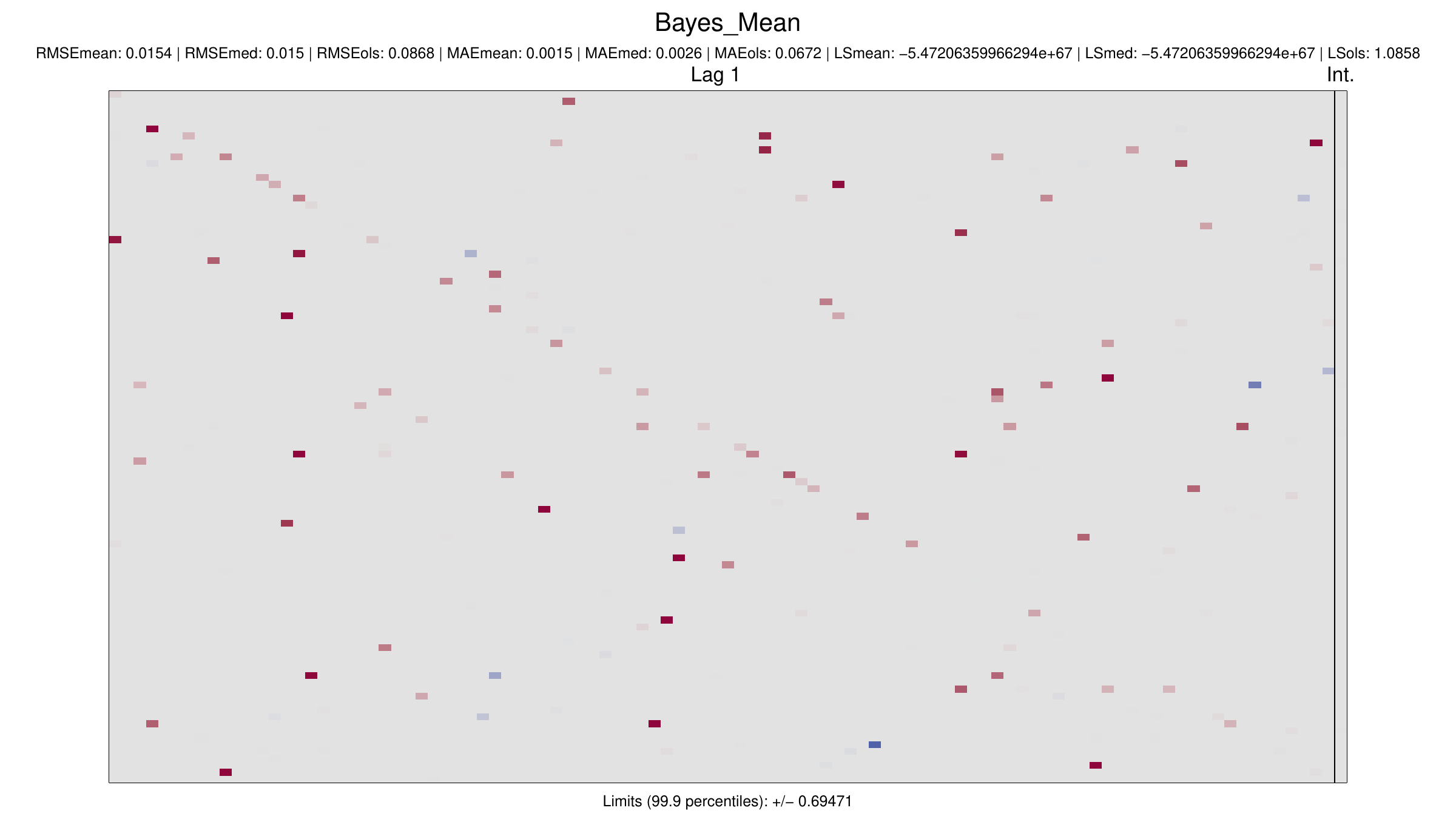} \\
 \includegraphics[page=1, width = .49\textwidth, trim = 60 0 52 70, clip = TRUE]{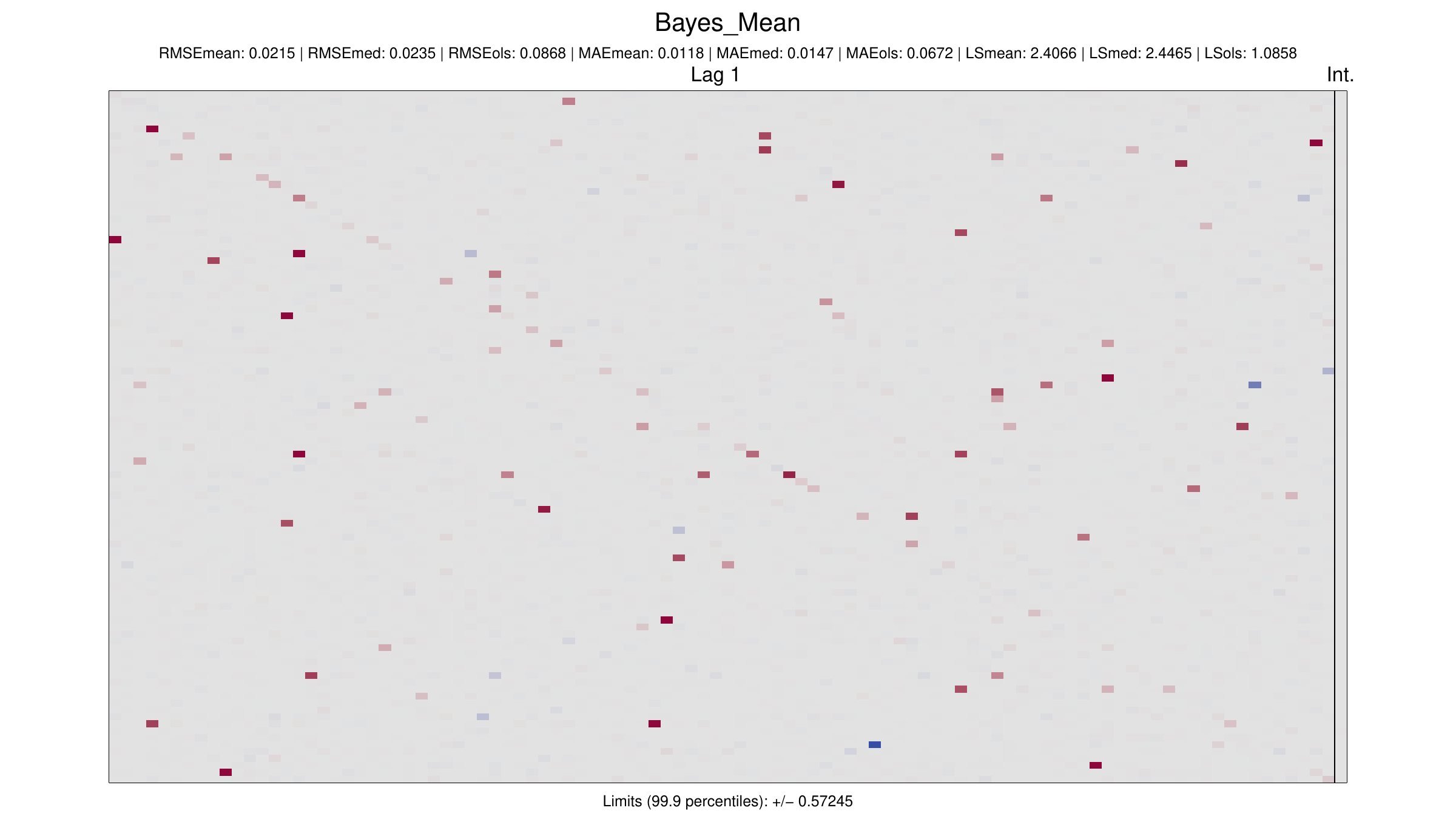}
 \includegraphics[page=1, width = .49\textwidth, trim = 60 0 52 70, clip = TRUE]{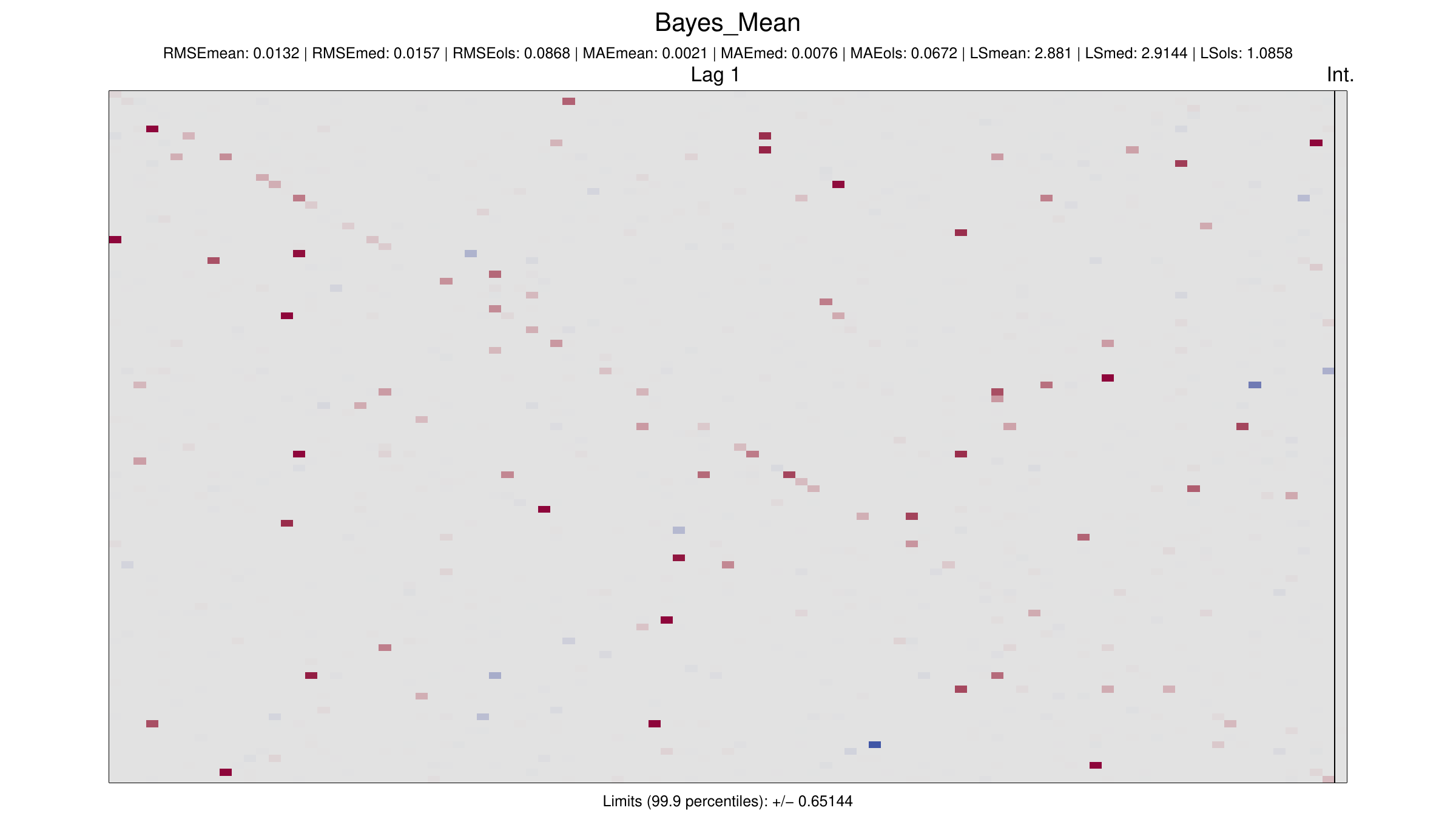} \\
 \includegraphics[page=1, width = .49\textwidth, trim = 60 0 52 70, clip = TRUE]{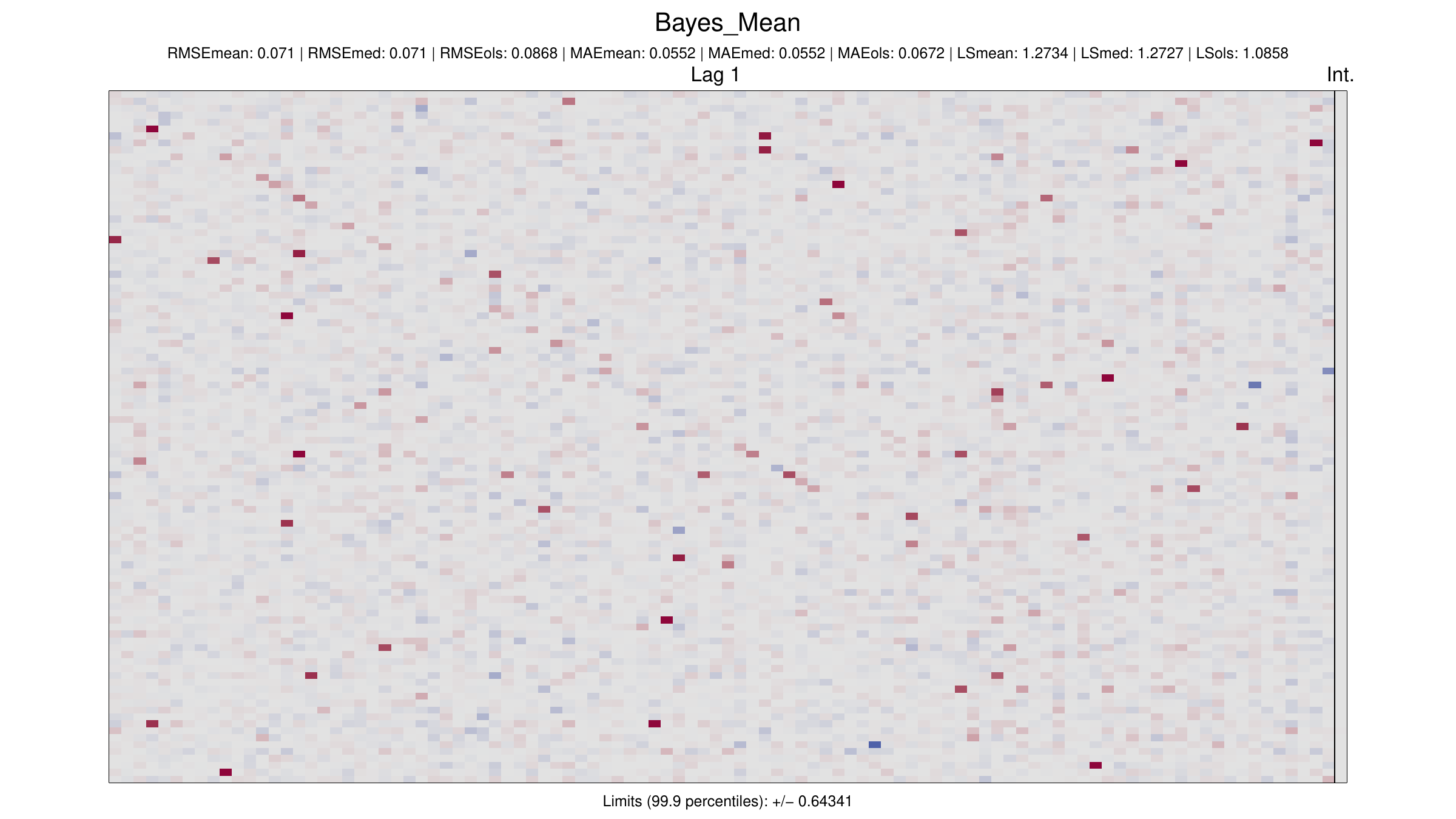}
 \includegraphics[page=1, width = .49\textwidth, trim = 60 0 52 70, clip = TRUE]{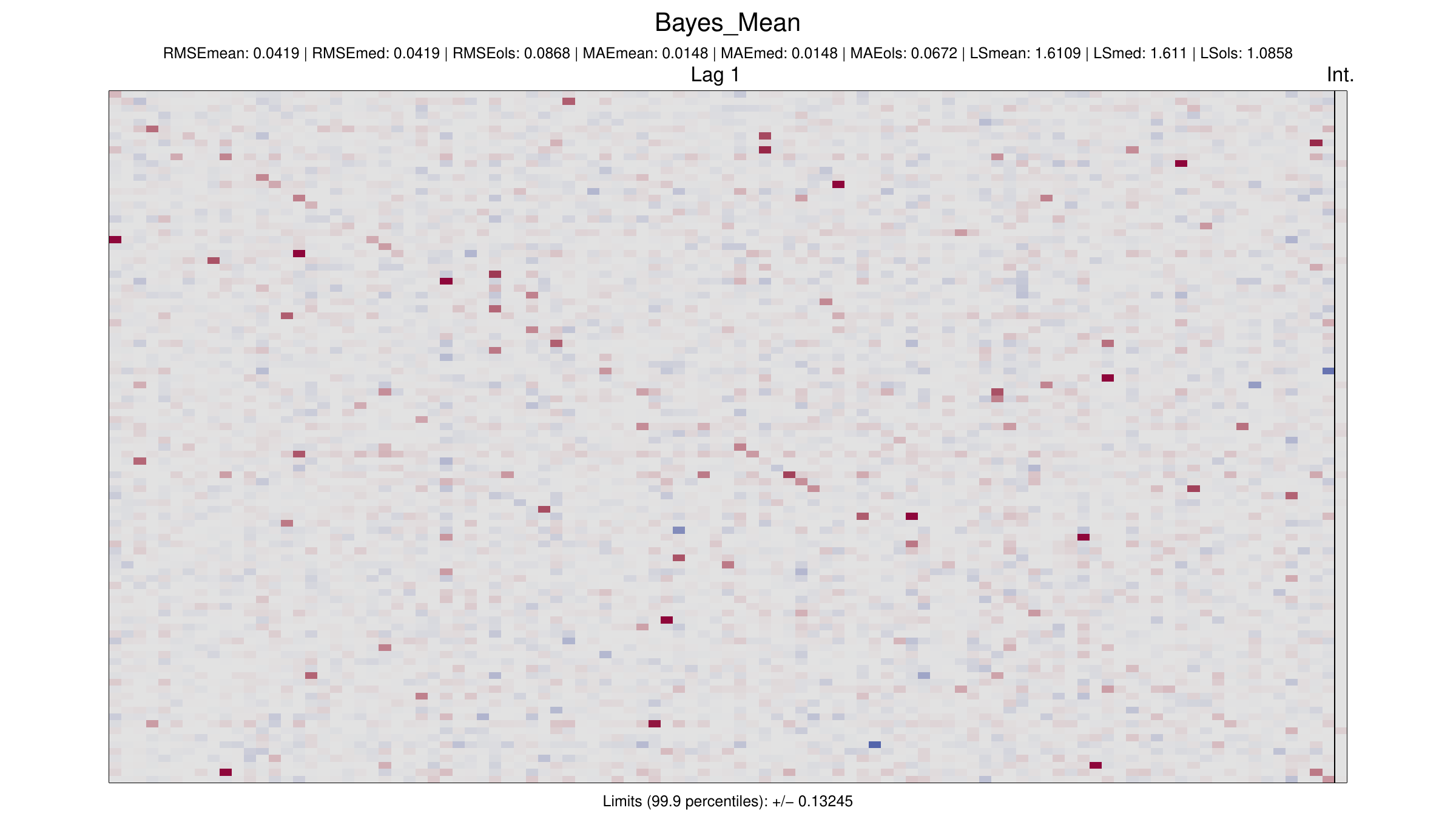}
 \caption{Exemplary visualization of the true and estimated VAR coefficients in the \emph{sparse} scenario where $T=250$ and $m=100$. Top left: DGP. Top right: OLS estimates. Second row: DL prior with $a_{DL} = 1/2$ (left) and $a_{DL} = 1/k = 1/101$ (right). Third row: NG prior with $a_{NG} = 1$ (left) and $a_{NG} = 1/10$ (right). Fourth row: Minnesota prior with $a_{M} = 1/1000$ (left) and $a_{M} = 1/10000$ (right).}
 \label{simstudyexample3}
\end{figure}

\begin{figure}
 \includegraphics[page=1, width = .49\textwidth, trim = 60 0 52 70, clip = TRUE]{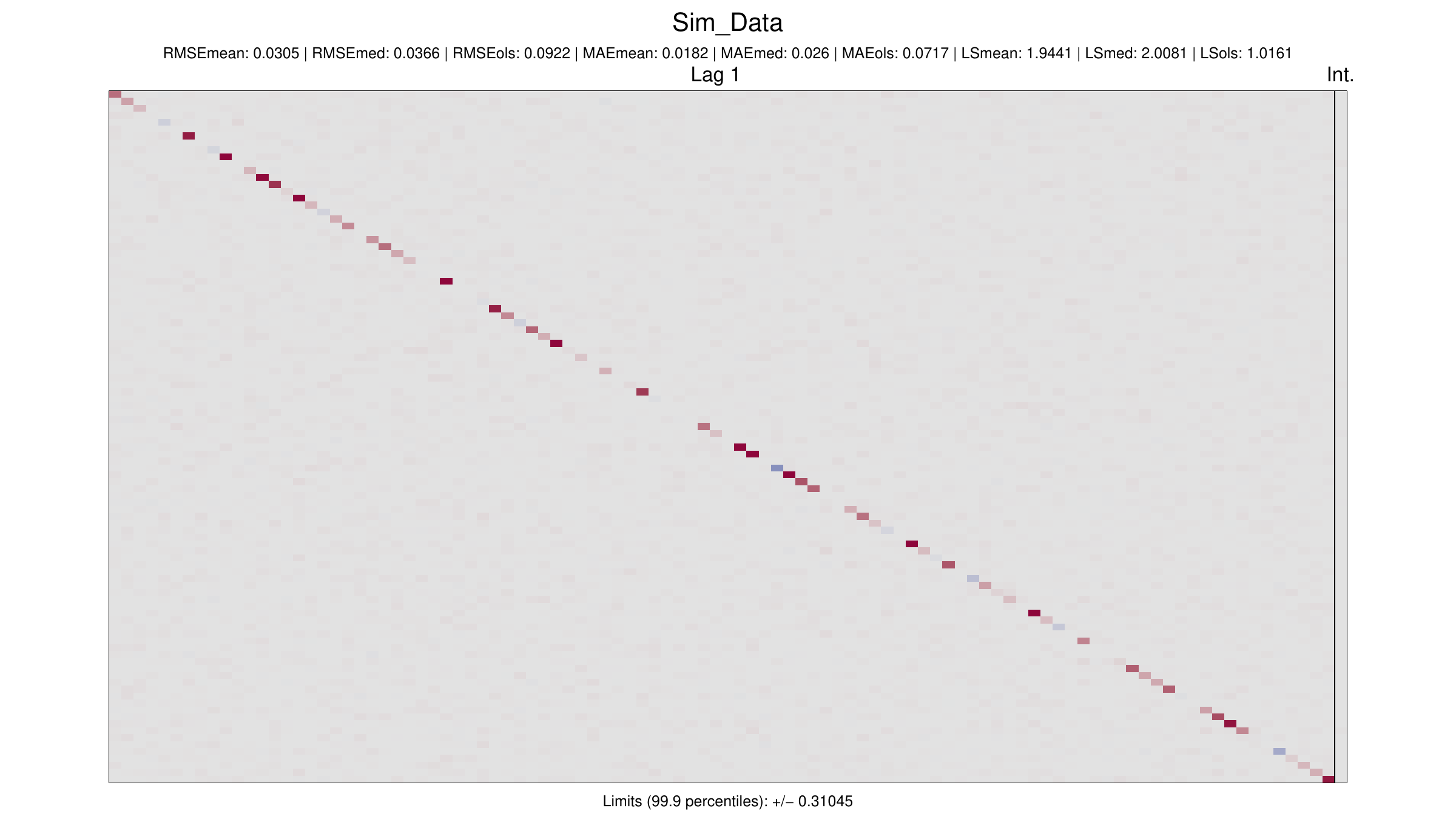}
 \includegraphics[page=4, width = .49\textwidth, trim = 60 0 52 70, clip = TRUE]{dense_T250_m100/DL2/001_heatmaps.pdf} \\
 \includegraphics[page=2, width = .49\textwidth, trim = 60 0 52 70, clip = TRUE]{dense_T250_m100/DL2/001_heatmaps.pdf}
 \includegraphics[page=1, width = .49\textwidth, trim = 60 0 52 70, clip = TRUE]{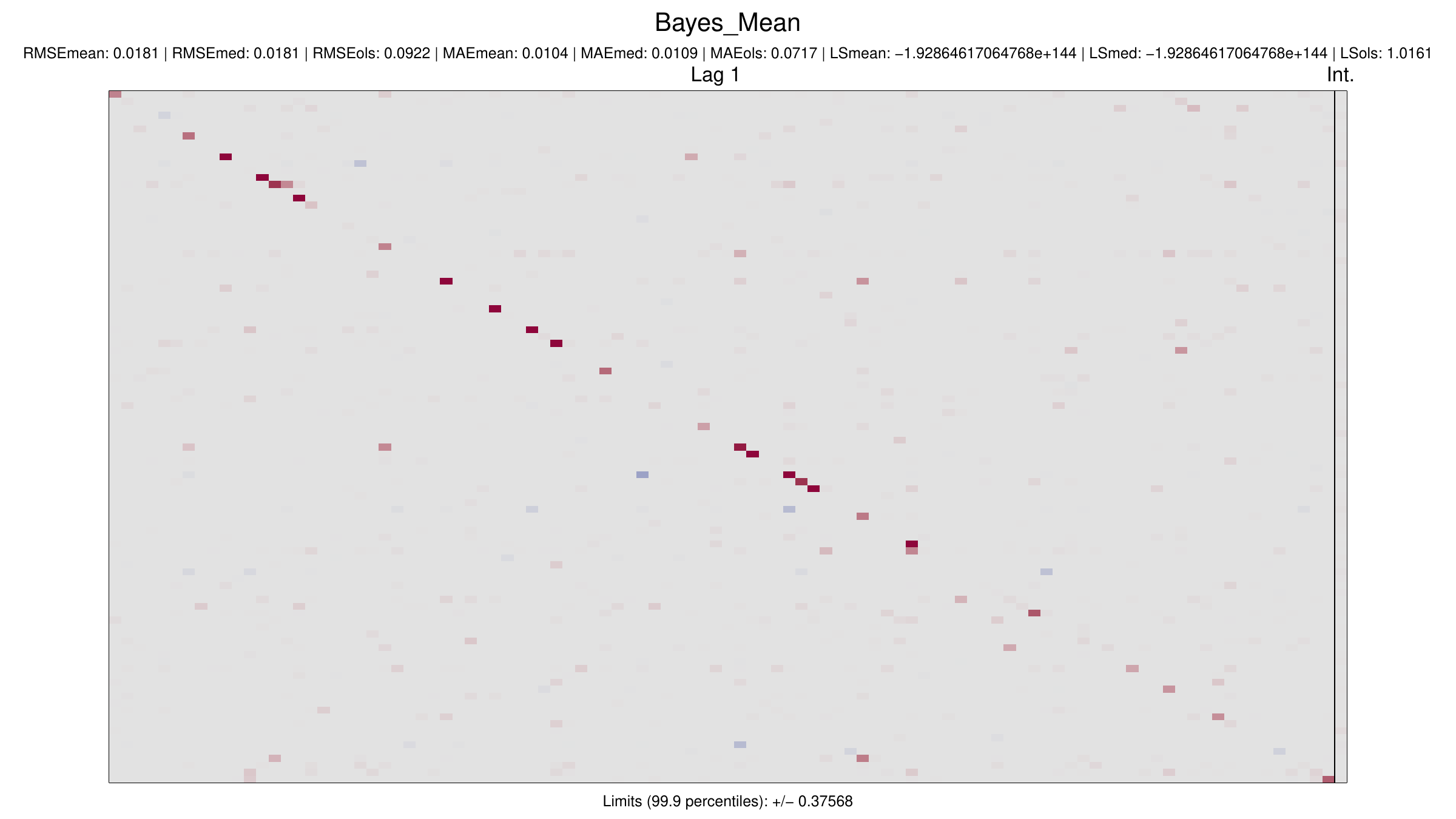} \\
 \includegraphics[page=1, width = .49\textwidth, trim = 60 0 52 70, clip = TRUE]{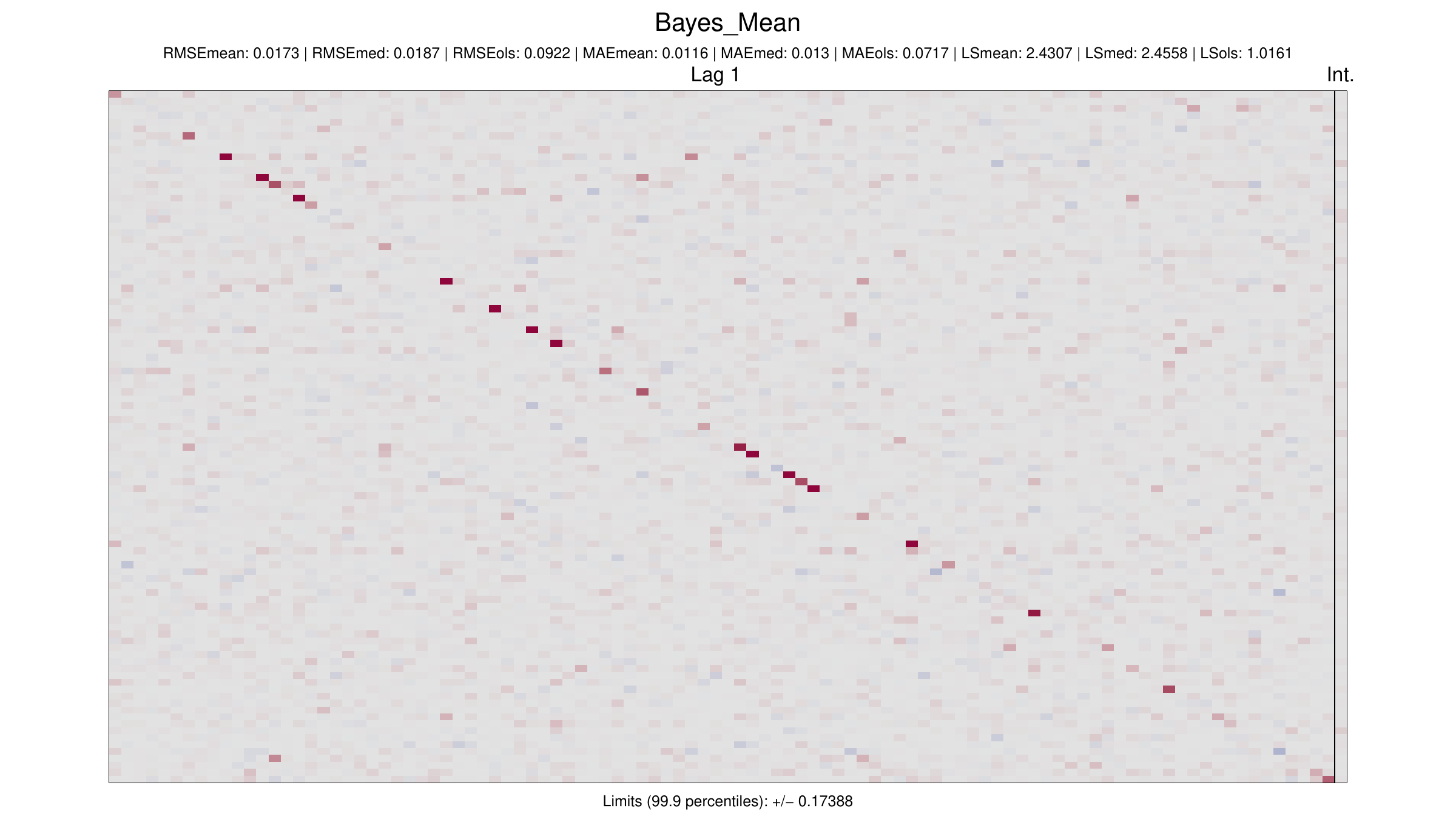}
 \includegraphics[page=1, width = .49\textwidth, trim = 60 0 52 70, clip = TRUE]{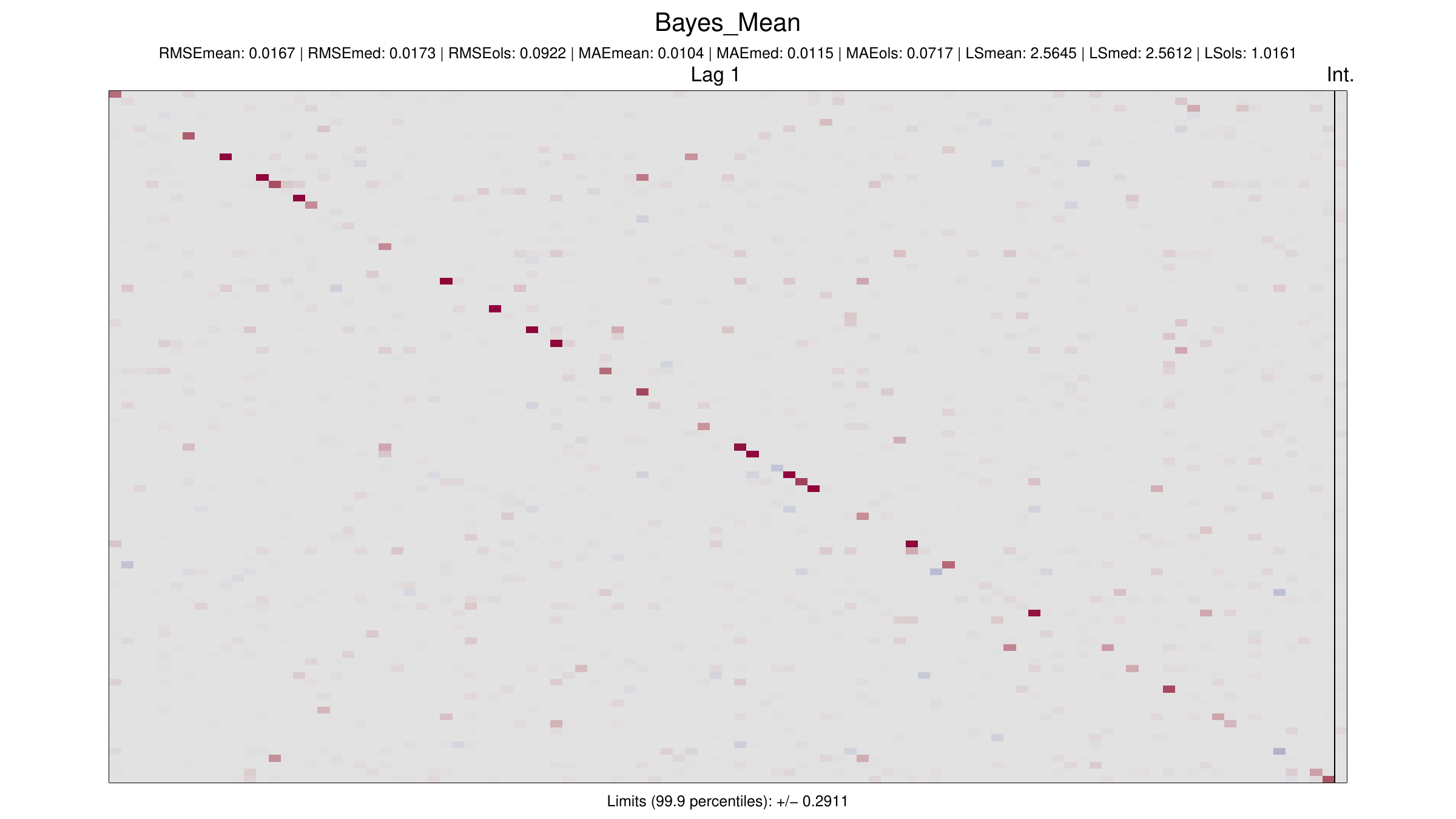} \\
 \includegraphics[page=1, width = .49\textwidth, trim = 60 0 52 70, clip = TRUE]{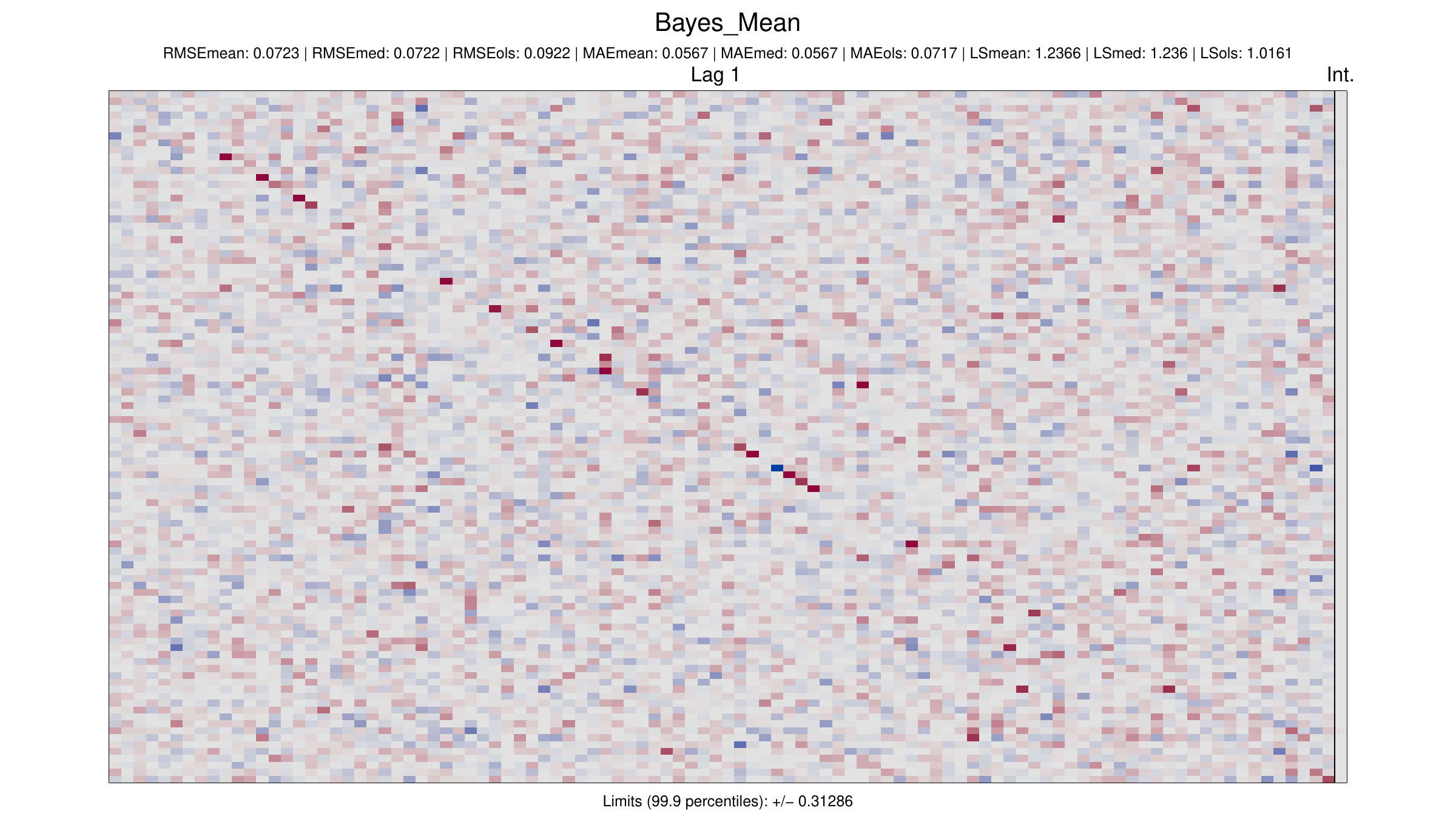}
 \includegraphics[page=1, width = .49\textwidth, trim = 60 0 52 70, clip = TRUE]{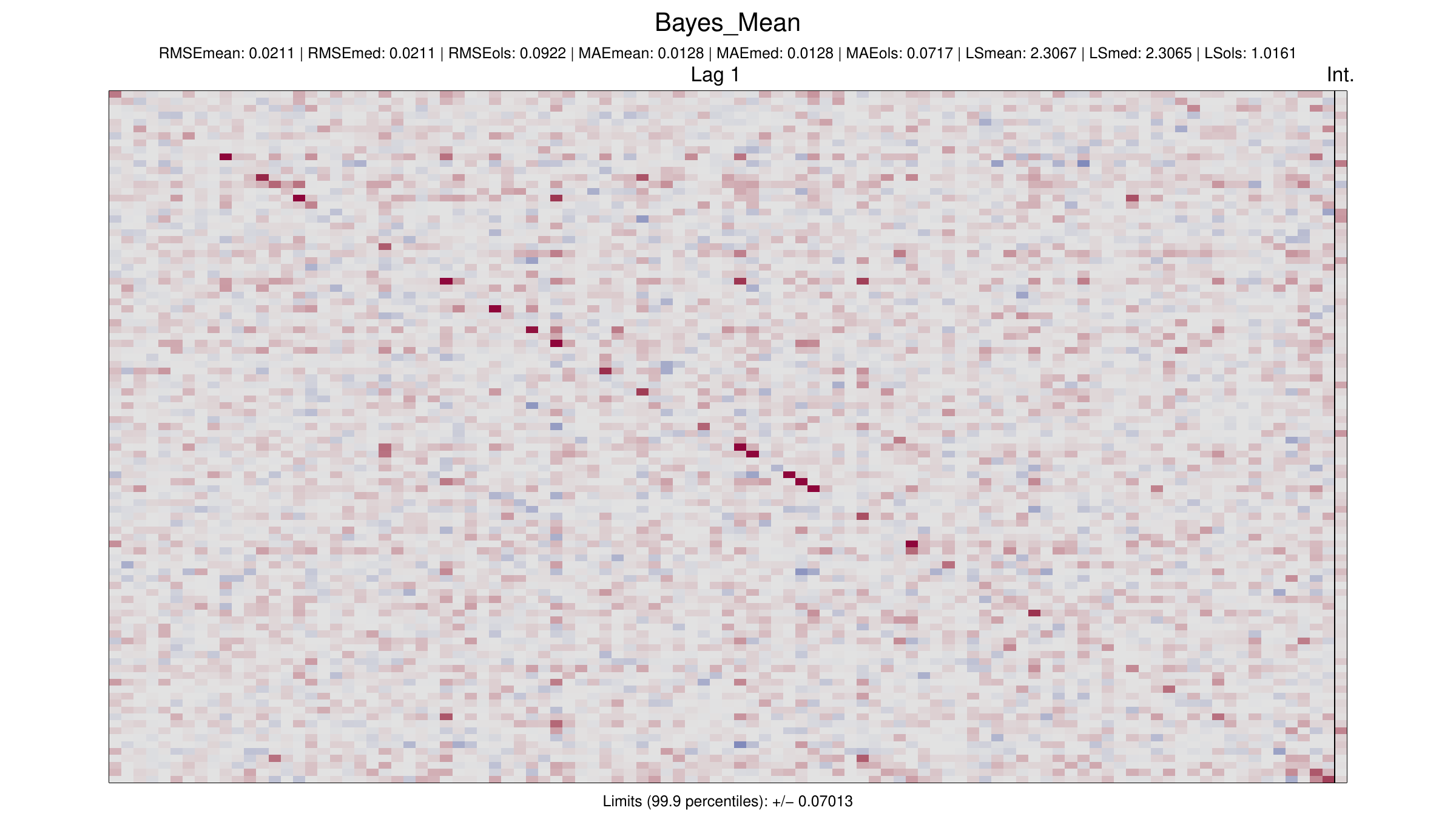}
 \caption{Exemplary visualization of the true and estimated VAR coefficients in the \emph{dense} scenario where $T=250$ and $m=100$. Top left: DGP. Top right: OLS estimates. Second row: DL prior with $a_{DL} = 1/2$ (left) and $a_{DL} = 1/k = 1/101$ (right). Third row: NG prior with $a_{NG} = 1$ (left) and $a_{NG} = 1/10$ (right). Fourth row: Minnesota prior with $a_{M} = 1/1000$ (left) and $a_{M} = 1/10000$ (right).}
 \label{simstudyexample4}
\end{figure}

First, we showcase four selected data generating scenarios (small + sparse, small + dense, large + sparse, large + dense) and visualize the posterior distribution of the VAR coefficients under seven different prior choices in Figures~\ref{simstudyexample1} to \ref{simstudyexample4}. For a comprehensive overview, see Table~\ref{tab:simstudy} in the main part of the paper.

\begin{figure}[p]
  \includegraphics[page=1, width=\textwidth, trim = 0 0 0 40, clip = TRUE]{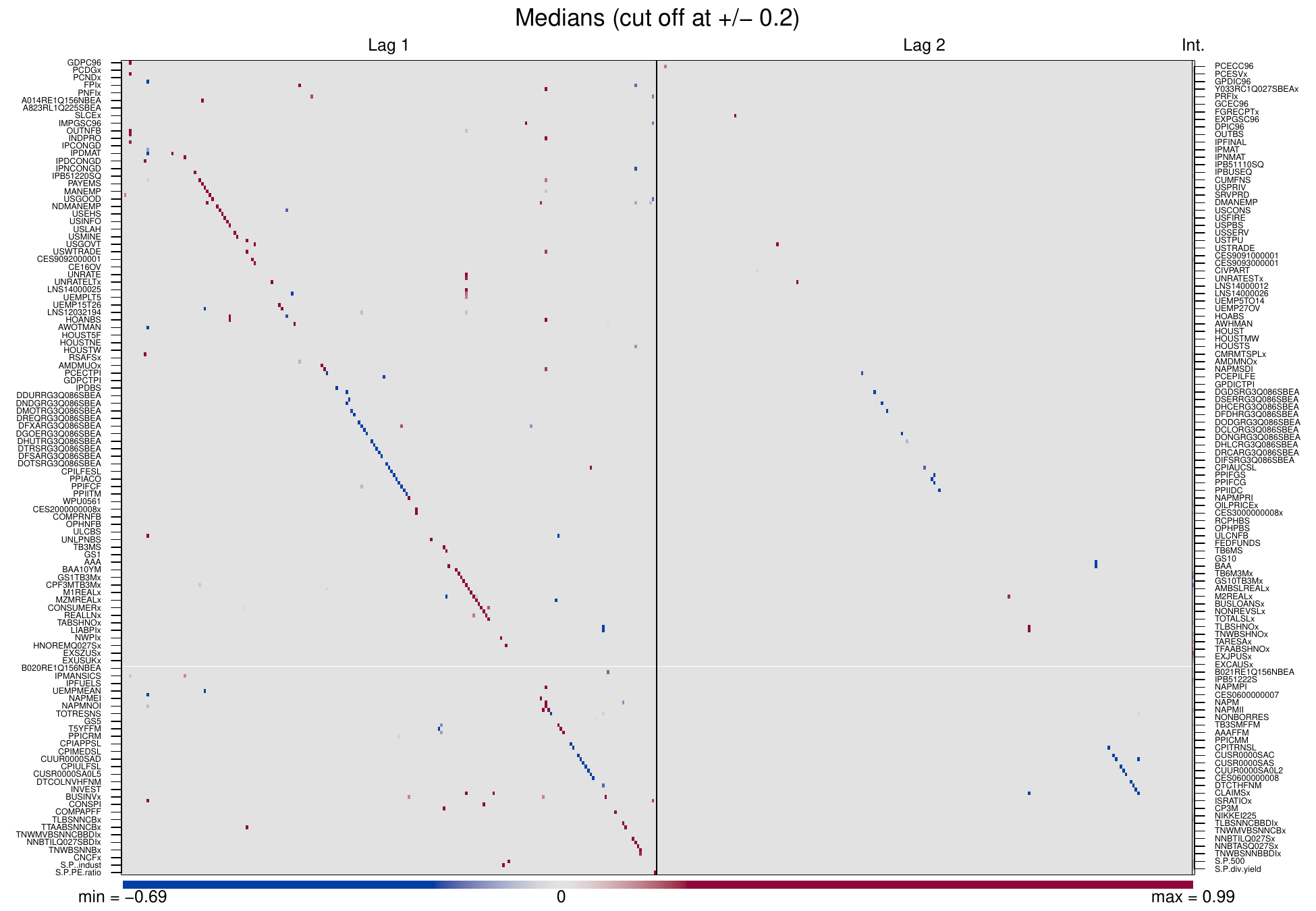}\\
  \includegraphics[page=1, width=\textwidth, trim = 0 0 0 40, clip = TRUE]{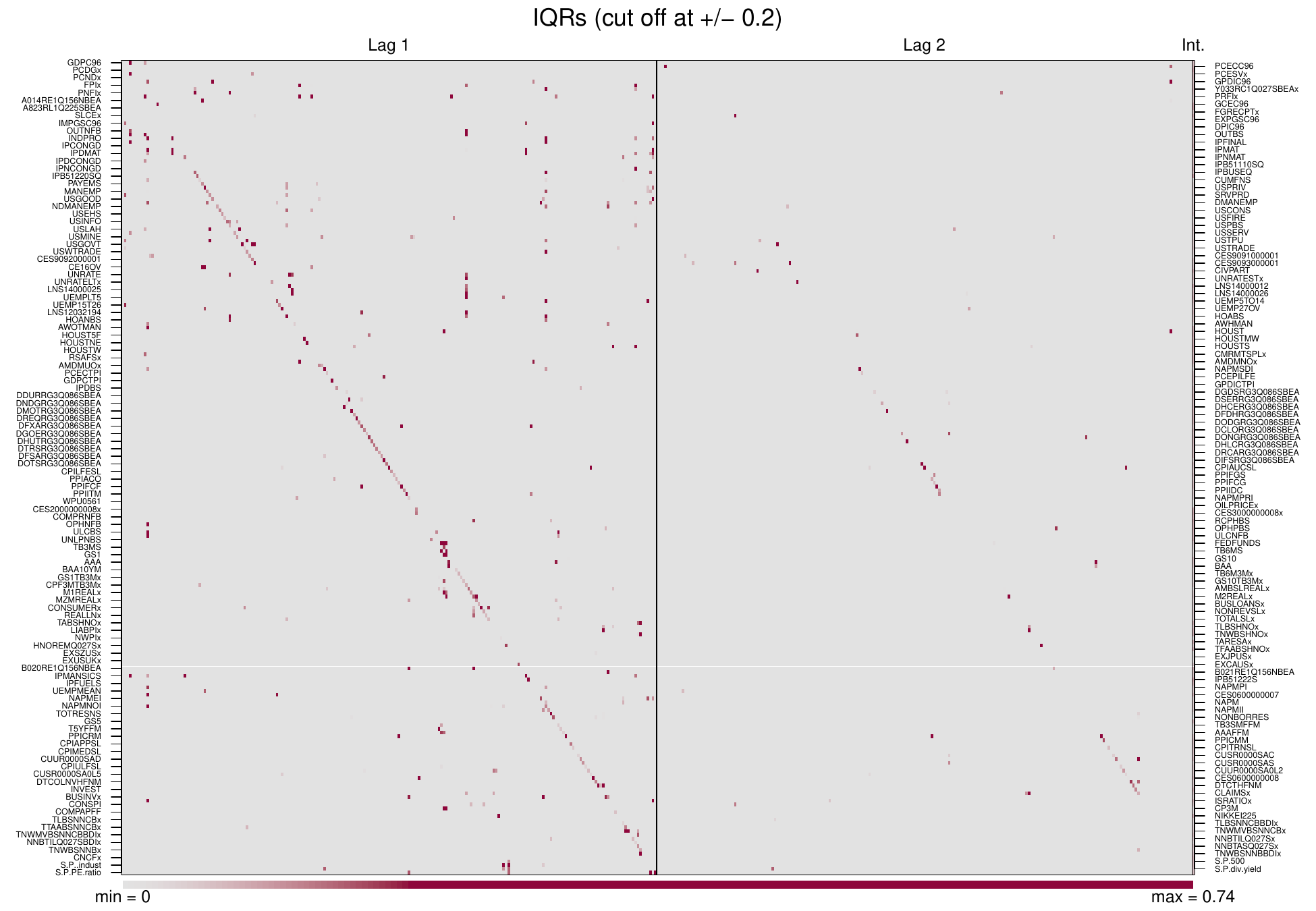}
\caption{Posterior medians (top) and posterior interquartile ranges (bottom) of VAR(2) coefficients, $a = 1/k = 1/431$.}
  \label{fig:ll2}
\end{figure}

\begin{sidewaysfigure}
  \centering
  \includegraphics[page=1, width=\textwidth, trim = 0 0 0 40, clip = TRUE]{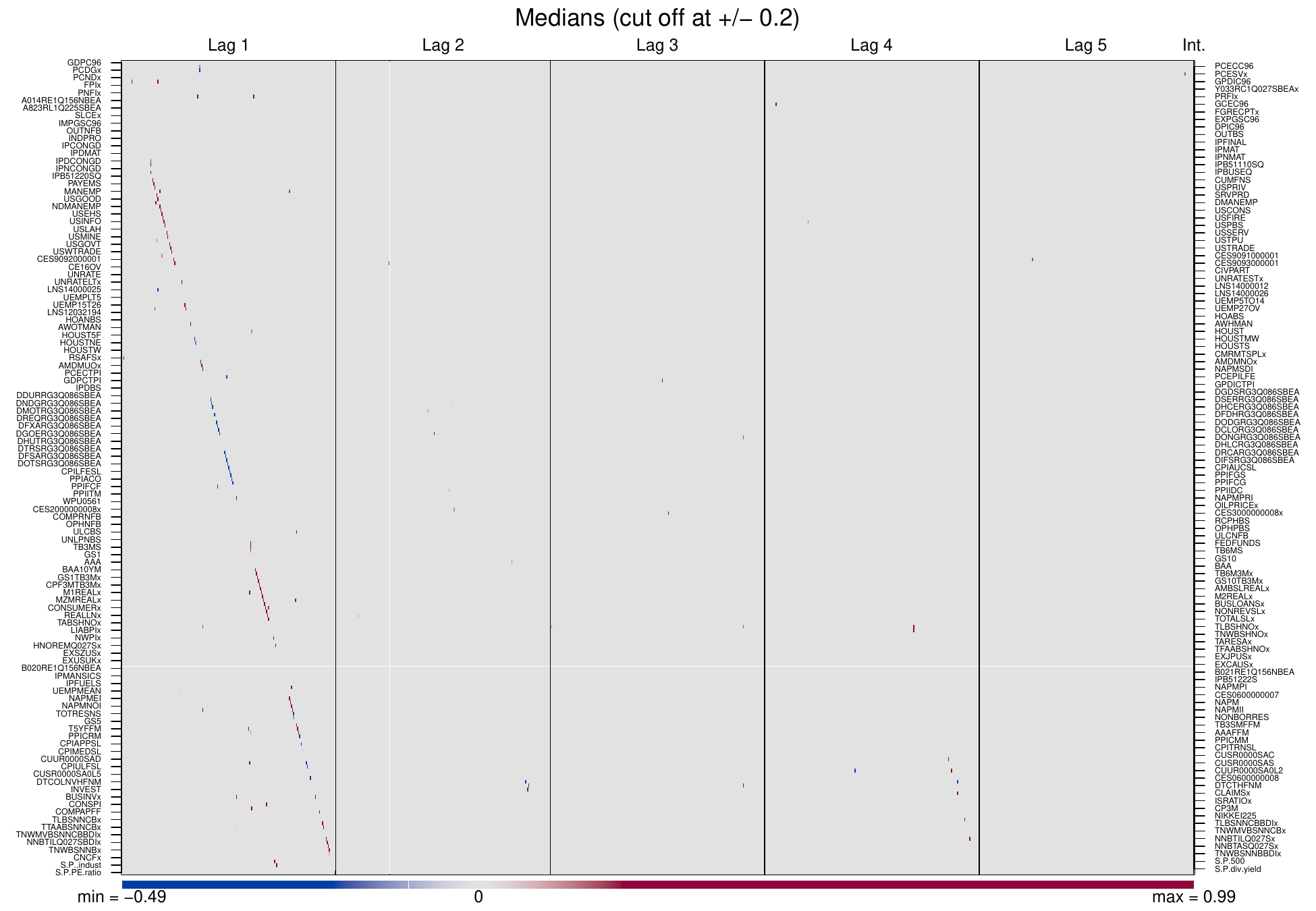}\\
\caption{Posterior medians of VAR(5) coefficients, $a = 1/k = 1/1076$.}
  \label{fig:ll5a}
\end{sidewaysfigure}

\begin{sidewaysfigure}
  \centering
  \includegraphics[page=1, width=\textwidth, trim = 0 0 0 40, clip = TRUE]{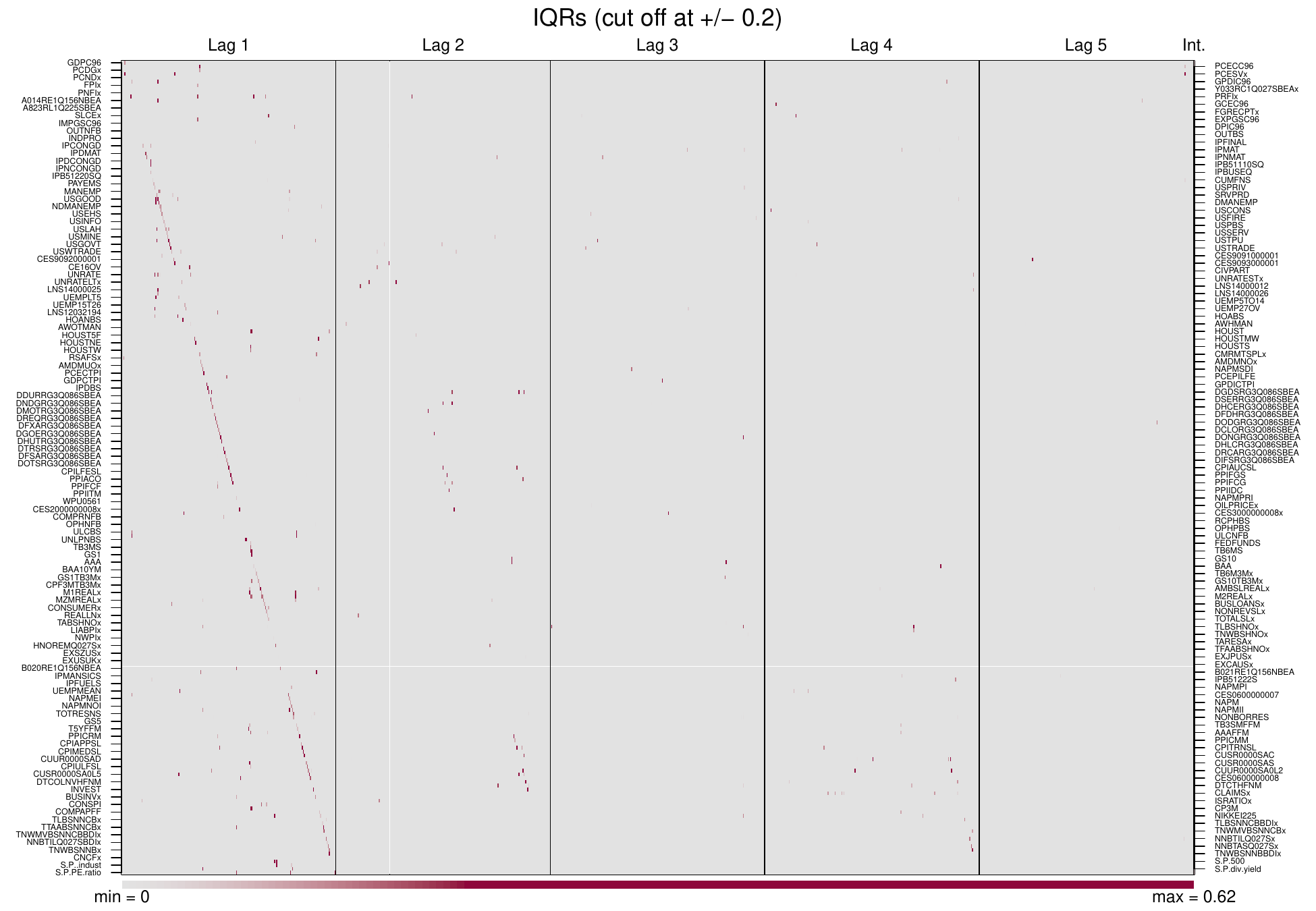}\\
\caption{Posterior interquartile ranges of VAR(5) coefficients, $a = 1/k = 1/1076$.}
  \label{fig:ll5b}
\end{sidewaysfigure}

Second, we illustrate results from a VAR(2)-FSV DL($1/k$) and a VAR(5)-FSV DL($1/k$) model for the US data in Figures~\ref{fig:ll2} to \ref{fig:ll5b}.

\end{appendix}

%\begin{supplement}
%\sname{Supplement A}\label{suppA} 
%\stitle{Title of the Supplement A}
%\slink[url]{http://www.some-url-address.org/dowload/0000.zip}
%\sdescription{Add description for supplement material.}
%\end{supplement}

%\begin{acknowledgement}
%And this is an acknowledgements section with a heading that was produced by the
%$\backslash$section* command. Thank you all for helping me writing this
%\LaTeX\ sample file. See \ref{suppA} for the supplementary material example.
%\end{acknowledgement}

\end{document}